\documentclass[
  ]
  {aipprocwjm}
\layoutstyle{6x9}
 
\usepackage{pennames}
\usepackage{graphicx}
\newcommand{\PZ}{\ensuremath{\mathrm{Z}}}
\renewcommand{\Pqb}{\ensuremath{\mathrm{b}}}
\renewcommand{\Paqb}{\ensuremath{\bar\mathrm{b}}}
\newcommand{\Pee}{\Pep\Pem}
\newcommand{\PWW}{\PWp\PWm}
\newcommand{\PeeWW}{\ensuremath{\Pee \kern -0.35em \rightarrow\PWW}}
\newcommand{\eV}{\ensuremath{\mathrm{e\kern-0.08em V}}}%
\newcommand{\MeV}{\ensuremath{\mathrm{M}{\eV}}}%
\newcommand{\GeV}{\ensuremath{\mathrm{G}{\eV}}}%
\newcommand{\LEP}{{\scshape lep}}
\newcommand{\ALEPH}{{\scshape aleph}}
\newcommand{\OPAL}{{\scshape opal}}
\newcommand{\DELPHI}{{\scshape delphi}}

\newcommand{\Lthree}{{\scshape l}{\small 3}}
\newcommand{\ZEUS}{{\scshape zeus}}

\newcommand{\etc}{{\it etc.}}%
\newcommand{\eg}{{\it e.g.}}%
\newcommand{\ie}{{\it i.e.}}%
\newcommand{\abs}[1]{\ensuremath{\left\vert{#1}\right\vert}}%
\newcommand{\dd}[1]{\ensuremath{\,{{\mathrm{d}}#1}}}%
\newcommand{\BEtt}{\ensuremath{{\mathrm{BE}_{32}}}}

\newcommand{\Qlong}{\ensuremath{Q_\mathrm{L}}}
\newcommand{\Qside}{\ensuremath{Q_\mathrm{side}}}
\newcommand{\Qout}{\ensuremath{Q_\mathrm{out}}}

\newcommand{\Rtrans}{\ensuremath{r_\mathrm{t}}}
\newcommand{\Rlong}{\ensuremath{r_\mathrm{L}}}
\newcommand{\Rside}{\ensuremath{r_\mathrm{side}}}
\newcommand{\Rout}{\ensuremath{r_\mathrm{out}}}

\newcommand{\deltaI}{\ensuremath{\delta_\mathrm{\,I}}}
\newcommand{\ycut}{\ensuremath{y_{\mathrm{cut}}}}

\begin{document}
 
\title{Bose-Einstein Correlations in e$^+$e$^-$ Annihilation and e$^+$e$^-\rightarrow \mathrm{W}^+\mathrm{W}^-$
\footnote{Talk given at {\it Fourth Workshop on Particle Correlations and Femtoscopy,}
          Krom\v{e}\v{r}\'{\i}\v{z}, Czech Republic, August 15--17, 2005.
          A shortened version of this report will appear in the proceedings.
          }
}
 
 
\author{W.~J.~Metzger}{
  address={Radboud University, Nijmegen, Netherlands}
}
 
%

\begin{abstract}
 Results on Bose-Einstein correlations in
 $\mathrm{e}^+\mathrm{e}^-\rightarrow\mathrm{hadrons}$
 are reviewed.
\end{abstract}
 
\maketitle
 
 
\section{Introduction}
 
  Bose-Einstein correlations (BEC) are those correlations which arise as a
  consequence of Bose symmetry, which leads to an enhancement of the production
  of identical particles close together in phase space.
  In this talk I will review some of the results on BEC from \Pee\ interactions.
 
  To study correlations among $q$ particles, we begin with the inclusive $q$-particle density,
\begin{equation} \label{eq:density}
   \rho_q(p_1,...,p_q) = \frac{1}{\sigma_\mathrm{tot}}
                         \frac{\mathrm{d}^q\sigma_q(p_1,...,p_q)}
                              {\mathrm{d}{p_1}...\mathrm{d}p_q}
\end{equation}
where $\sigma_q$ is the inclusive $q$-particle cross section.
These densities are normalized such that
\begin{eqnarray}
    \int\rho_1(p)\dd{p} &=& \langle n\rangle  \\
    \int\rho_2(p_1,p_2)\dd{p_1}\dd{p_2} &=& \langle n(n-1)\rangle \\
    \int\rho_2(p_1,p_2,p_3)\dd{p_1}\dd{p_2}\dd{p_3} &=& \langle n(n-1)(n-2)\rangle \ ,
\end{eqnarray}
\etc, and are related to the factorial cumulants, $C$, by
\begin{eqnarray}
           \rho_1(p_1) &=& C_1(p_1) \\
           \rho_2(p_1,p_2) &=& C_1(p_1)C_1(p_2)+C_2(p_1,p_2)           \\
           \rho_3(p_1,p_2,p_3)) &=& C_1(p_1)C_1(p_2)C_1(p_3) \nonumber \\
                                &+& \sum_{\textrm{3\ perms}}C_1(p_1)C_2(p_2,p_3) \nonumber \\
                                &+& C_3(p_1,p_2,p_3) \ .
\end{eqnarray}
The $q$-particle correlations are then measured by
$\rho_q - \prod_{i=1}^q C_1(p_i1)$.
For $q>2$, the correlations are a sum of
``trivial'', \ie, arising trivially from correlations of
smaller $q$, and ``genuine'' correlations.
For example, for $q=3$
the trivial correlations are given by $\sum_{\textrm{3\ perms}}C_1(p_1)C_2(p_2,p_3)$
and the genuine correlations by $C_3$.
 
It is convenient to normalize $\rho$ and $C$:
\begin{eqnarray}
   \mathsf{R}_q  &=&  \frac{\rho_q}{\prod_{i=1}^q \rho_1(p_i)} \label{eq:Rqorig}  \\
          {K}_q  &=&  \frac{C_q}{\prod_{i=1}^q \rho_1(p_i)}    \label{eq:Kqorig} \ .
\end{eqnarray}
 
  In order to study BEC, and not other correlations, one usually studies the ratio of the above-defined
  $\mathsf{R_q}$ to the $\mathsf{R_q}$ that one expects in the absence of BEC.
This is equivalent to replacing
the product of single-particle densities in (\ref{eq:Rqorig}) by the $q$-particle density  expected when BEC
is absent, $\rho_{0q}$, giving:
\begin{equation}
   R_q   =   \frac{\rho_q}{\rho_{0q}} \ . \label{eq:Rq}
\end{equation}
  This ratio is usually regarded as
a function of $Q$, where $Q^2=M_q^2-(qm)^2$ with $M_q$ the mass of the $q$ particles
and $m$ the mass of each particle.
  If the particles have identical 4-momenta. $Q=0$.
  For 2 particles, $Q^2$ is simply the 4-momentum difference.
  Thus, \eg, 2-particle BEC are studied using
\begin{equation} \label{Rtwo_def}
  R_2(Q)= \frac{\rho(Q)}{\rho_0(Q)} \ ,
\end{equation}
where the subscript $q$ has been suppressed.
 
It can be shown in a variety of ways that $R_q$ is related to the spatial distribution of
the particle production \cite{GGLP:1960,Boal:1990}.
For example, assuming incoherent particle production and a spatial source density of pion emitters, $S(x)$,
leads to
\begin{equation} \label{Rtwo_F}
  R_2(Q)= 1 + \abs{G(q)}^2 \ .
\end{equation}
where $G(Q)\!=\!\int\!\dd{x}\,e^{\imath Qx} S(x)$ is the Fourier transform of $S(x)$.
Assuming $S(x)$ is a Gaussian with radius $r$ results in
\begin{equation} \label{Rtwo_Gauss}
  R_2(Q)= 1 + e^{-Q^2r^2} \ .
\end{equation}
 
Customarily, an additional parameter, $\lambda$, is introduced in (\ref{Rtwo_Gauss}):
\begin{equation} \label{Rtwo_G}
  R_2(Q)= 1 + \lambda e^{-Q^2r^2} \ .
\end{equation}
This parameter is meant to account  for several effects:
\begin{itemize}
  \item partial coherence.  Completely coherent particle production would imply $\lambda=0$.
  \item multiple sources.
  \item particle purity, \eg, experimental difficulty in distinguishing pions from kaons.
\end{itemize}
 
The assumption of a spherical (radius $r$) Gaussian distribution of particle emitters
seems unlikely in \Pee\ annihilation, where there is a definite jet structure.
However, we must keep in mind that the BEC only occur among particles produced close to each other in phase
space.  In a two-jet event, a particle produced in the core of one jet would not be ``close'' to a particle
produced in the core of the other jet.
The volume in which hadronization occurs may thus be larger than a sphere of radius $r$ and is not
necessarily spherical.

Furthermore, no time dependence has been considered, \ie, the source has been assumed to be static, which is
certainly wrong.
 
A number of other parametrizations have been considered in the literature.
Nevertheless, in spite of the above-noted limitations, this Gaussian parametrization (\ref{Rtwo_G}) is the
one most frequently used by experimentalists.
When this Gaussian parametrization does not fit well, an expansion about the Gaussian
(Edgeworth expansion \cite{Edgeworth}) can be used instead. Keeping only the lowest-order non-Gaussian term,
$\exp(-Q^2r^2)$ becomes $\exp(-Q^2r^2)\cdot\left[1+\frac{\kappa}{3!}H_3(Qr)\right]$, where $H_3$ is the
third-order Hermite polynomial.
 
In the interest of comparison of as much data as possible, I shall only consider results using the
Gaussian or Edgeworth parametrizations here.

\section{Experimental Difficulties}
 
There are a number of experimental problems which affect the results on BEC and their interpretation.
 
Particle purity influences the value of $\lambda$.  In studying BEC of pions, it is often
assumed that all particles are pions, which is not true.  For example, in \PZ-boson decays approximately
15\% of the particles are not pions.  This lowers the observed value of $\lambda$.
The value of $r$ for BEC involving a particle from a long-lived resonance and one produced directly,
or involving particles from different long-lived resonances,
will be larger than for two particles produced directly.  Consequently, the resulting enhancement at small
$Q$ in $R_2(Q)$ will be narrower, possibly narrower than the experimental resolution, which in turn
reduces the observed value of $\lambda$.  On the other hand, the effect of short-lived resonances is
to increase the observed value of $r$, since it takes into account, in some average way, the distance the
resonance travels before decaying.  In \PZ-boson decays only about 16\% of charged pions are produced
directly, while 62\% come from short-lived ($\Gamma>6.7\,\MeV$) and 22\% from long-lived resonances.
Particles from weak decays are produced far away from the others, so that there are no BEC between pions
from the weak decay and the rest. This results in a smaller value of $\lambda$. About 20\% of \PZ-bosons
decay to \Pqb\Paqb.
Not only the values of $\lambda$ and $r$ may be influenced by resonances, but also the shape of $R_2$.
One can attempt to correct for all of these effects, typically by using a Monte Carlo model.
However, the correctness of the model is open to question.
 
\begin{figure}
  \includegraphics*[width=.5\textwidth]{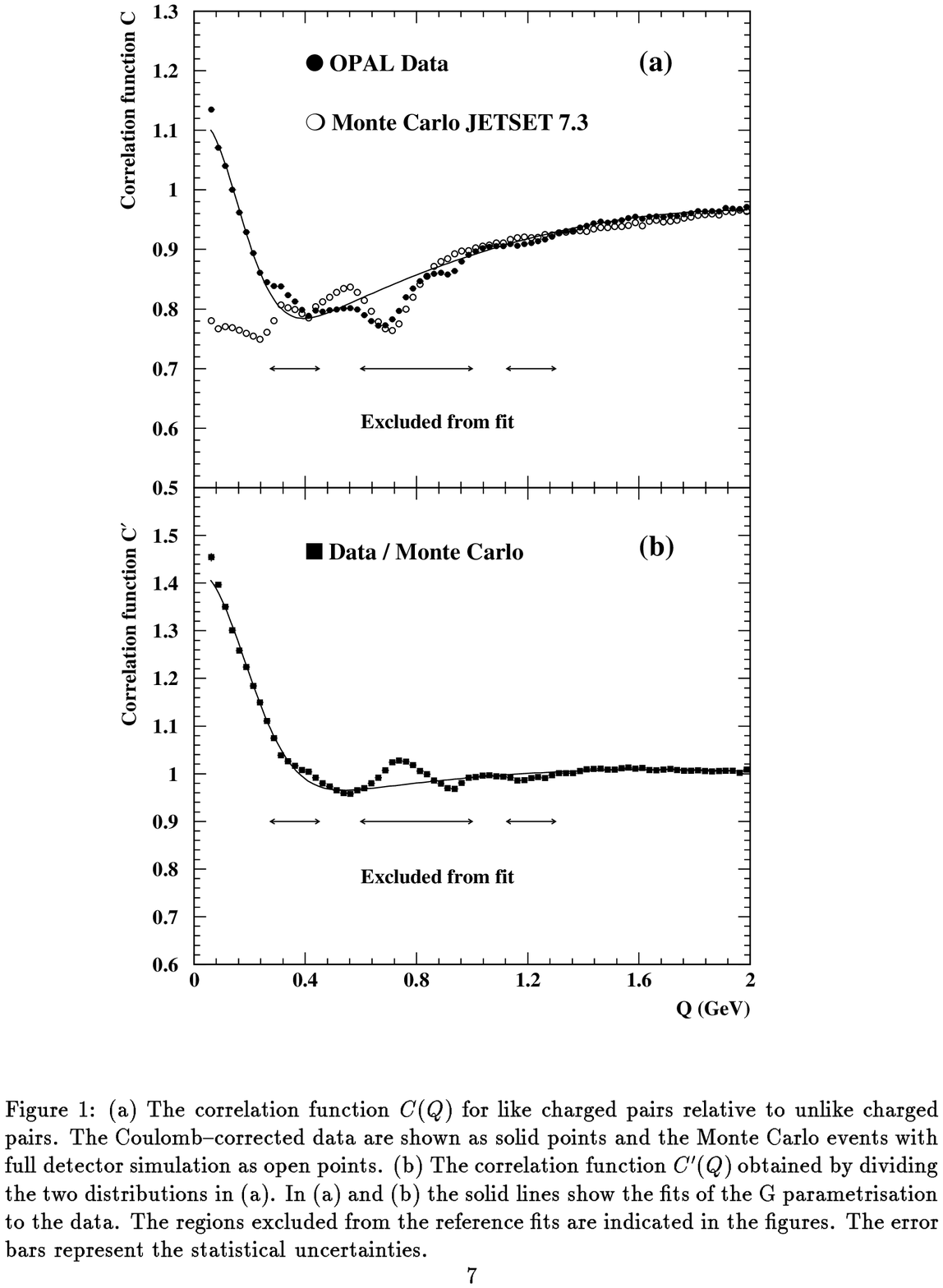}
  \caption{(a) $R_2$ (here called $C$) for \PZ-decay data and Monte Carlo;
           (b) the ratio of $R_2$ of the data to that of the Monte Carlo.
          }
  \label{fig:unlikepairs}
\end{figure}

A second problem is the choice of the so-called reference sample, \ie, the sample for which $\rho_0$ is the
density.  Since it is impossible to turn off Bose statistics, this sample does not exist.  It must be
created artificially.  Common choices are unlike-sign pairs, Monte Carlo models, and mixed events.
When studying BEC in like-sign pion pairs one can use unlike-sign pion pairs to form the reference sample.
The main problem is that the unlike-sign pairs have resonances which the like-sign pairs do not.  This is
illustrated in Fig.~\ref{fig:unlikepairs}a, where $R_2$ calculated using an unlike-sign pair reference
sample is shown for \PZ-decay data and for Monte Carlo.  Structure in $R_2$ due to resonances in the
reference sample is clearly seen.  One might think of correcting for this using Monte Carlo, but this will only work
if the resonances are well-described in the Monte Carlo.  This is clearly not the case here, as is seen in
Fig.~\ref{fig:unlikepairs}b.  The solution adopted is to exclude the affected regions from the fit of
(\ref{Rtwo_G}) to the data.
These regions, indicated in Fig.~\ref{fig:unlikepairs}, cover large ranges in $Q$, and one must ask whether
larger, or additional, excluded regions are needed.
 
Using Monte Carlo as the reference sample avoids the different resonances of like- and unlike-sign pairs,
but is still not completely free of resonance problems.  Reflections of unlike-sign resonances and
multi-body resonance decays, \eg, $\Pghpr\rightarrow\Pgpp\Pgpm\Pgh\rightarrow\Pgpp\Pgpm\Pgpp\Pgpm\Pgpz$,
still affect $\rho_0$.  But these effects are small compared to the use of unlike-sign pairs as the
reference sample.   Another problem with using a Monte Carlo reference sample is the description of
fragmentation.  In a Monte Carlo model, fragmentation is controled by some parameters whose values are
chosen to give a good description of the data.  However, the data contain BEC.  If the Monte Carlo does not
include a description of BEC, the fragmentation parameters will be chosen so as to describe (in so far as is
possible) BEC as well as the fragmentation itself.  The reference sample will then have some enhancement in
$\rho$ at small $Q$, and its use as reference sample will result in a smaller value of $\lambda$.  The value
of $r$ may also be affected.  On the other hand, a BEC model can be included in the Monte Carlo and the
fragmentation parameters determined.  Then the BEC model can be turned off to construct the reference
sample.  How good this reference sample is depends on how well the BEC were modeled.  To model them well
would require not only knowing how to correctly model BEC but also knowing the results of the BEC
analysis before doing it.
 
Event mixing involves taking pairs of particles from different events.  For each real event, a similar event
is constructed by replacing each particle by a particle from a different event.  This procedure, of course,
removes all correlations, not just BEC.  Therefore, care should be taken in
choosing the particles in order to have a reference sample similar to the real sample in characteristics
such as jet structure. In this way some kinematic correlations can be preserved.  Other correlations can be
reinstated using the results of applying the same mixing procedure to Monte Carlo, \ie, by using the double
ratio  $R_2/R_2^\mathrm{MC}$ instead of $R_2$ itself.
Thus this method is also not entirely free of Monte Carlo uncertainties.
For 2-jet events one can mix hemispheres instead of events.  In this approach the reference sample is
constructed by pairing a particle from one hemisphere with a particle from the other hemisphere which has
been reflected through the origin.

A third problem is final-state interactions, both Coulomb and strong.
Identical charged particles will be repulsed by the Coulomb interaction, decreasing the observed value of
$\lambda$ and increasing the observed value of $r$.
The most common way to take this into account is to multiply $R_q$ by a correction factor before fitting
(\ref{Rtwo_G}) to the data.  The correction factor most commonly used is the so-called Gamow factor
\cite{Gyulassy:1979}, which is the modulus square of the non-relativistic Coulomb wave function at the
origin.  For 2-particle BEC, given the resolution of the \LEP\ experiments, the correction is only a few per
cent in the lowest $Q$ bin (twice that if an unlike-sign reference sample is used), and is therefore
frequently ignored.  However for 3-particle BEC it is of the order of 10\%.
 
\begin{figure}
  \hfill  \includegraphics*[width=0.90\textwidth]{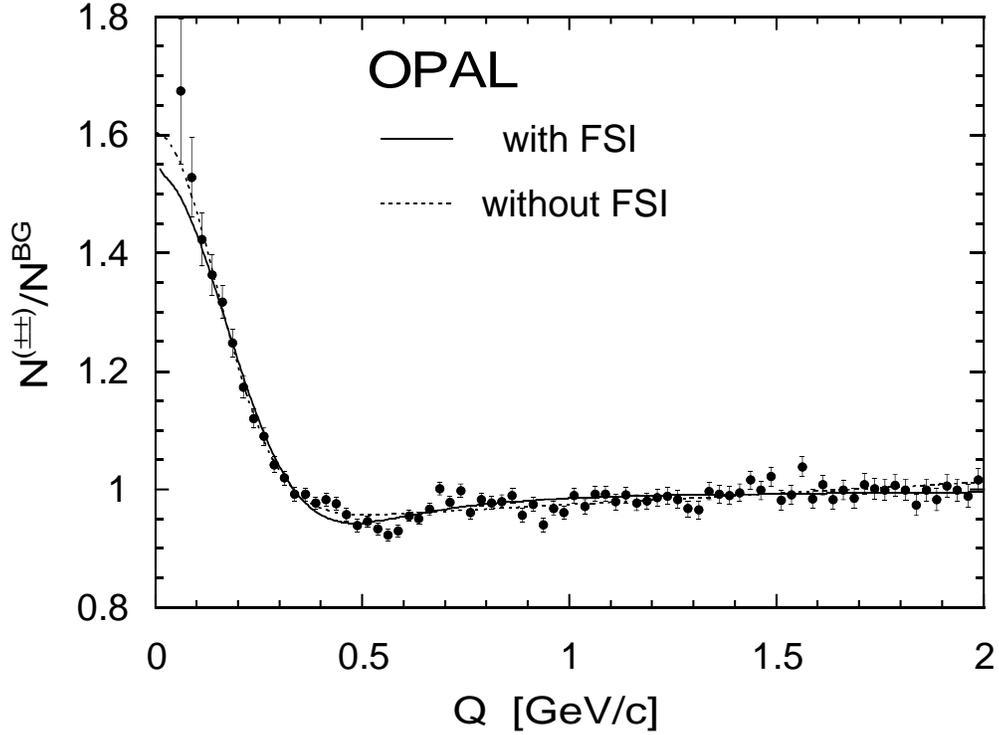}
  \caption{Fits with and without FSI (Coulomb and $S=0$ $\pi\pi$ phase shifts) to \OPAL\ Z-decay data \cite{Osada:1996}.}
  \label{fig:opal_fsi}
\end{figure}

A different approach is to incorporate both the Coulomb and the $S=0$ $\pi\pi$ phase shifts into the wave
function and derive a corrected formula for $R_2$.  An example of this approach is
shown in Fig.~\ref{fig:opal_fsi} \cite{Osada:1996}.
Significant differences are found in the values of~$\lambda$ and~$r$, which are listed in Table~\ref{tab:fsi}.
However, such sophisticated approaches have not been used by any of the experimental groups.

\begin{table}
\begin{tabular}{lcc}
\hline
      {FSI}          &  {$\lambda$}      &    {$r$} (fm)         \\
\hline
     with            &  $1.04\pm0.03$    & $1.09\pm0.04$  \\
     without         &  $0.71\pm0.04$    & $1.34\pm0.04$  \\
\hline
\end{tabular}
 \caption{Results of fits \cite{Osada:1996}
 with and without FSI (Coulomb and $S=0$ $\pi\pi$ phase shifts) to \OPAL\ Z-decay data.}
\label{tab:fsi}
\end{table}
 
Finally, there is the effect of long-range correlations not adequately taken into account by the reference
sample.  As observed, \eg, in Fig.~\ref{fig:opal_fsi}, $R_2$ is not constant at large $Q$.  To account for
this in fitting the data, the right hand side of (\ref{Rtwo_G}) is multiplied by an appropriate factor,
usually a linear dependence on $Q$,
\begin{equation} \label{eq:Rtwo_Glin}
  R_2(Q)= \gamma \left(1 + \lambda e^{-Q^2r^2}\right) \left(\vphantom{e^{Q^2}}1+\delta Q\right) \ ,
\end{equation}
although a quadratic dependence may also be used if necessary,
and the normalization, $\gamma$, is usually left as a free parameter.

\section{Experimental Results}
Now let us turn to a comparison of results from various \Pee\ experiments.
 
\subsection{2-particle BEC}
\begin{figure}
 \rotatebox{270}{
  \includegraphics*[width=.6\textwidth,bbllx=86,bblly=128,bburx=519,bbury=772]{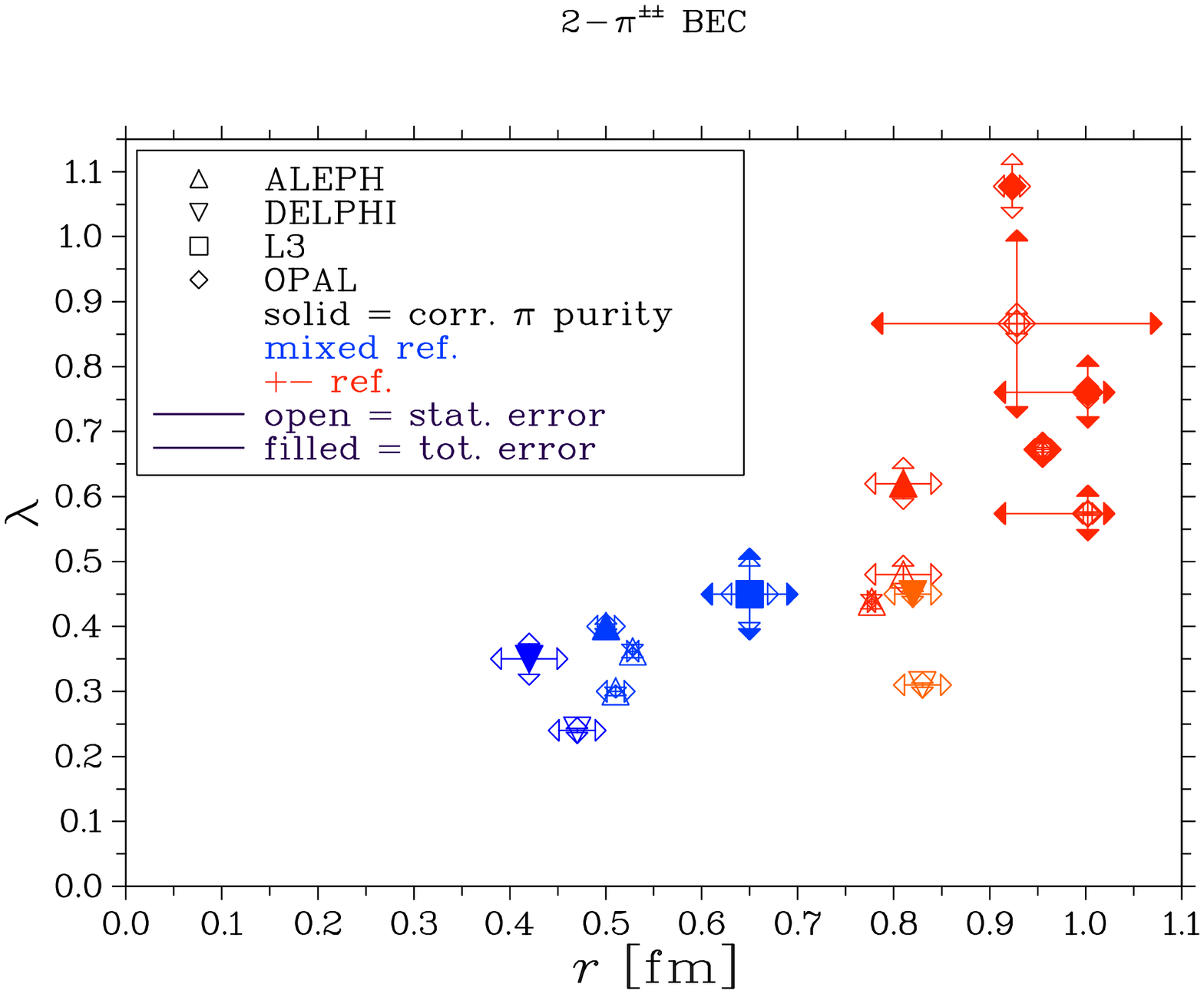}
 }
  \caption{$\lambda$ and $r$ at $\sqrt{s}=M_\mathrm{Z}$ found in the \LEP\ experiments
   \cite{ALEPH:1992,ALEPH:2004,DELPHI:1992,L3_3pi:2002,OPAL:1991,OPALmult:1996,OPAL3D:2000}.
   Open arrow-heads at the end of error bars indicate that the uncertainties are statistical only,
   filled arrow-heads that the uncertainties are combined statistical and systematic.
   }
  \label{fig:lep}
\end{figure}

\paragraph{Dependence on the reference sample}
The values of $\lambda$ and $r$ found using identical charged-pion pairs from hadronic \PZ\ decays
in the \LEP\ experiments, \ALEPH\ \cite{ALEPH:1992,ALEPH:2004},
\DELPHI\ \cite{DELPHI:1992}, \Lthree\ \cite{L3_3pi:2002} and
\OPAL\ \cite{OPAL:1991,OPALmult:1996,OPAL3D:2000} are displayed in Fig.~\ref{fig:lep}.
 
Although these points are all supposed to be measurements of the same quantitities and are all
determined by fitting the Gaussian parametrization to $R_2$,
there is little agreement among the points.
Clearly there must be large systematic uncertainties not accounted for.
Indeed, most of the points have only statistical error bars.
However, some have error bars including also systematic uncertainties.
But these are clearly insufficient.
 
Solid points are corrected for pion purity, while open points are not.
It is apparent that this correction increases the value of $\lambda$
but has little effect on the value of $r$.
 
All of the results with $r>0.1$ fm were obtained using an unlike-sign reference sample,
while all of those with smaller $r$ were obtained with a mixed reference sample.
It is clear that the choice of reference sample has a large effect on the values of the parameters.
In comparing the results of different experiments, we must therefore be sure that the reference samples used
are comparable.

\begin{figure}
 \rotatebox{270}{
  \includegraphics*[width=.6\textwidth,bbllx=86,bblly=128,bburx=519,bbury=772]{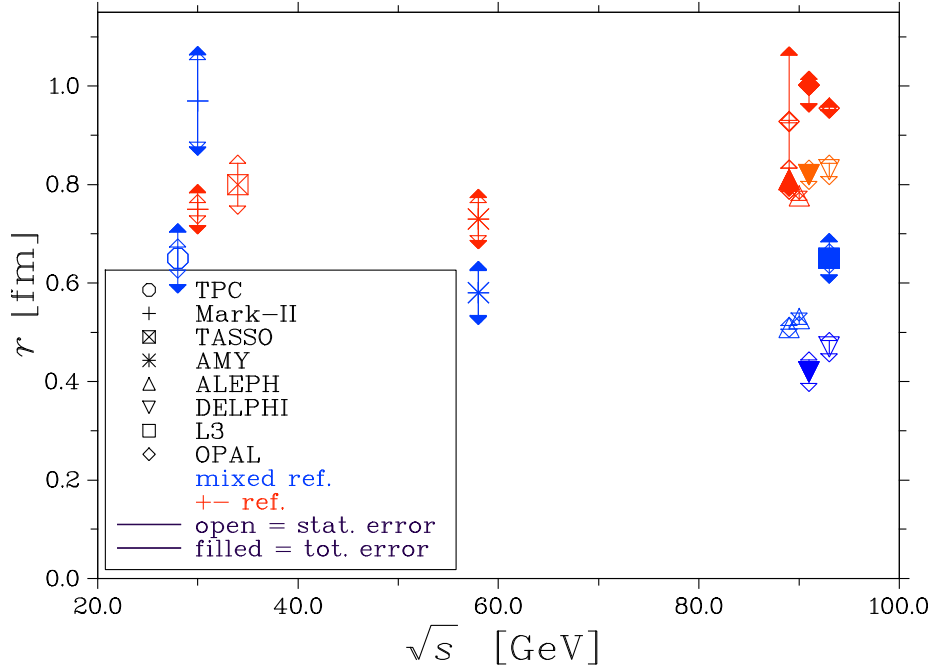}
 }
  \caption{$\sqrt{s}$ dependence of $r$
\cite{TPC:1985,MarkII:1989,TASSO:1986,AMY:1995,ALEPH:1992,ALEPH:2004,DELPHI:1992,L3_3pi:2002,OPAL:1991,OPALmult:1996,OPAL3D:2000}.
           For clarity some points are shifted slightly in $\sqrt{s}$.
   Open arrow-heads at the end of error bars indicate that the uncertainties are statistical only,
   filled arrow-heads that the uncertainties are combined statistical and systematic.
           }
  \label{fig:roots}
\end{figure}

\paragraph{Dependence on the center-of-mass energy}
The values of $r$ found using identical charged-pion pairs in \Pee\ annihilation is shown {\it versus} the
center of mass energy, $\sqrt{s}$, in Fig.~\ref{fig:roots}.  Keeping in mind that we should compare only
results using the same reference sample, we conclude that there is no evidence for a $\sqrt{s}$ dependence.
 
\begin{figure}
 \rotatebox{270}{
  \includegraphics*[width=.6\textwidth,bbllx=86,bblly=128,bburx=519,bbury=772]{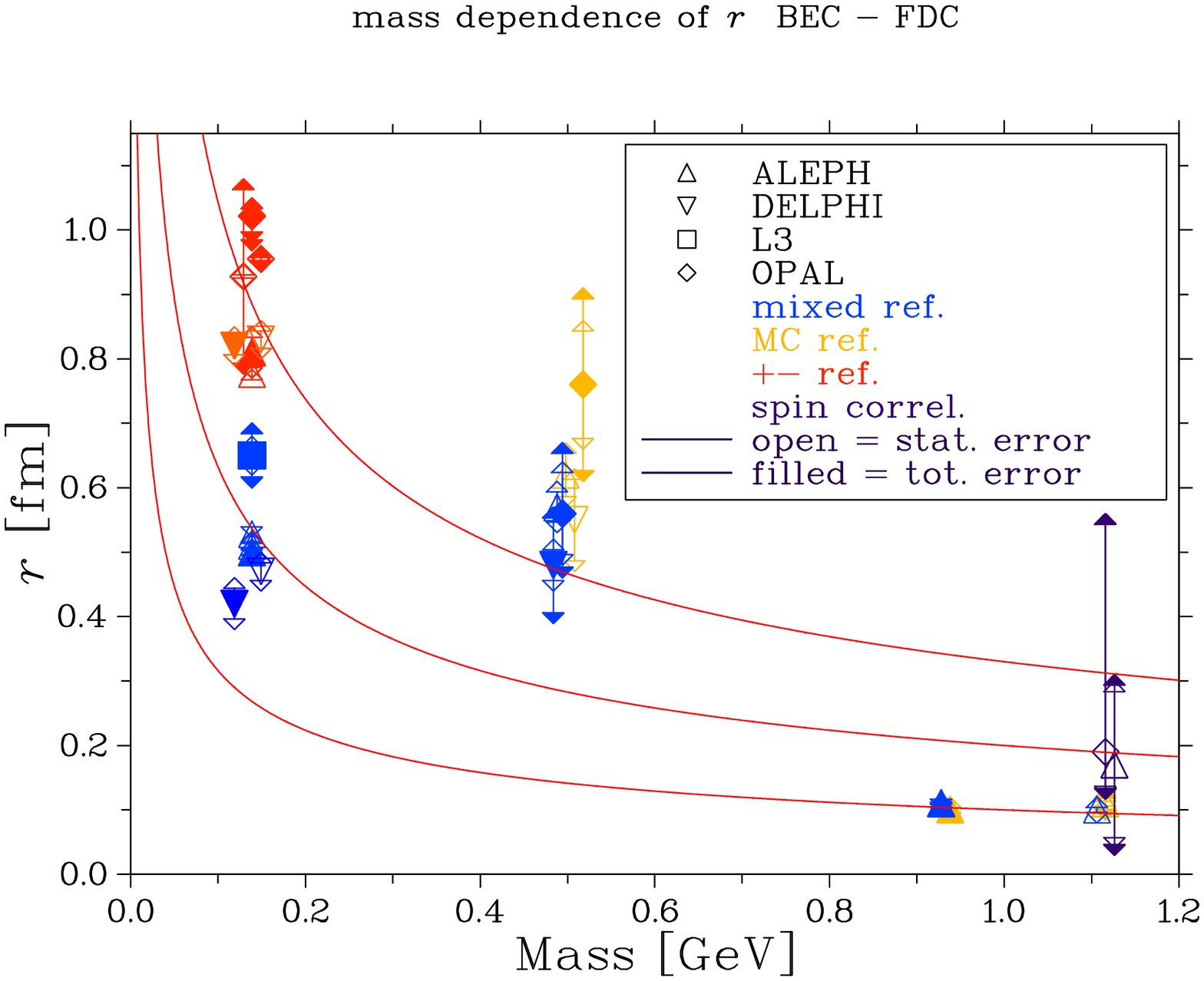}
 }
  \caption{Dependence of $r$ on the mass of the particle as determined at $\sqrt{s}=M_\mathrm{Z}$  from
           2-particle BEC for charged pions
           \cite{ALEPH:1992,ALEPH:2004,DELPHI:1992,L3_3pi:2002,OPAL:1991,OPALmult:1996,OPAL3D:2000},
           charged kaons \cite{DELPHIK:1996,OPALKp:2001} and
           neutral kaons \cite{ALEPHnon_pi:2005,DELPHIK:1996,OPALK0:1995}
           and from Fermi-Dirac correlations for protons \cite{ALEPHnon_pi:2005}
           and lambdas \cite{ALEPHlam:2000,OPALlam:1996}.
   Open arrow-heads at the end of error bars indicate that the uncertainties are statistical only,
   filled arrow-heads that the uncertainties are combined statistical and systematic.
           The curves illustrate a $1/\sqrt{m}$ dependence.
          }
  \label{fig:r-m}
\end{figure}

\paragraph{Dependence on the particle mass}
It has been suggested, on several grounds \cite{Alexander:2003}, that $r$ should depend on the mass of the
particle as $r\propto1/\sqrt{m}$.  Results from \LEP\ experiments on $r$ from 2-particle BEC for charged
pions and for kaons, as well as the corresponding Fermi-Dirac correlations for protons and lambdas, are
shown in Fig.~\ref{fig:r-m}.  Again, restricting our comparison to results which have used the same type of
reference sample (in this case mixed), we see no evidence for a $1/\sqrt{m}$ dependence.  Rather, the data
suggest one value of $r$ for mesons and a smaller value for baryons.  The value for baryons, about 0.1 fm,
seems very small; if true it is telling us something unexpected about the mechanism of baryon production.

\begin{figure}
   \includegraphics*[width=0.32\textwidth,bbllx=101,bblly=167,bburx=447,bbury=710]{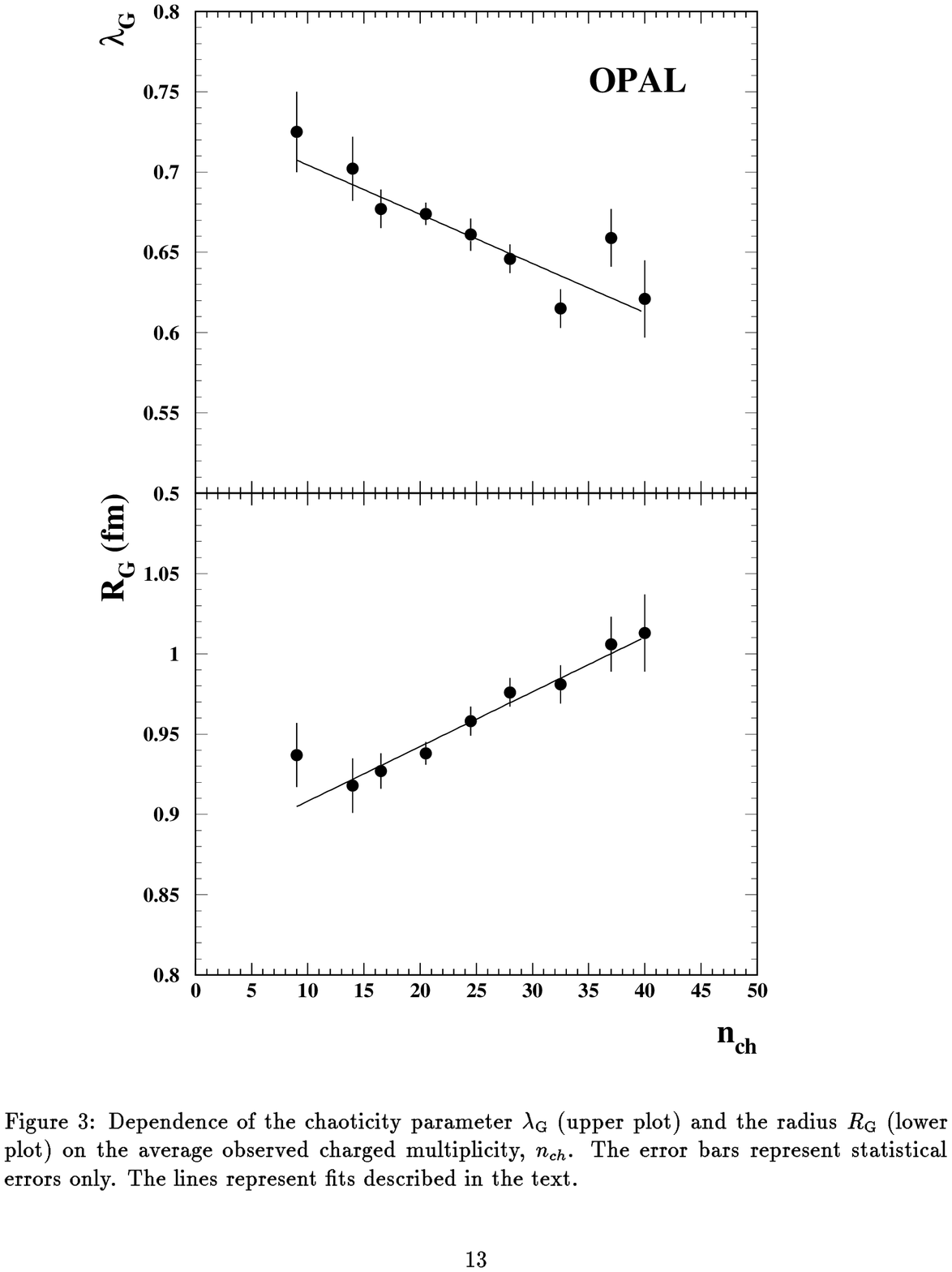}
  \hfil
   \includegraphics*[width=0.32\textwidth,bbllx=100,bblly=180,bburx=449,bbury=723]{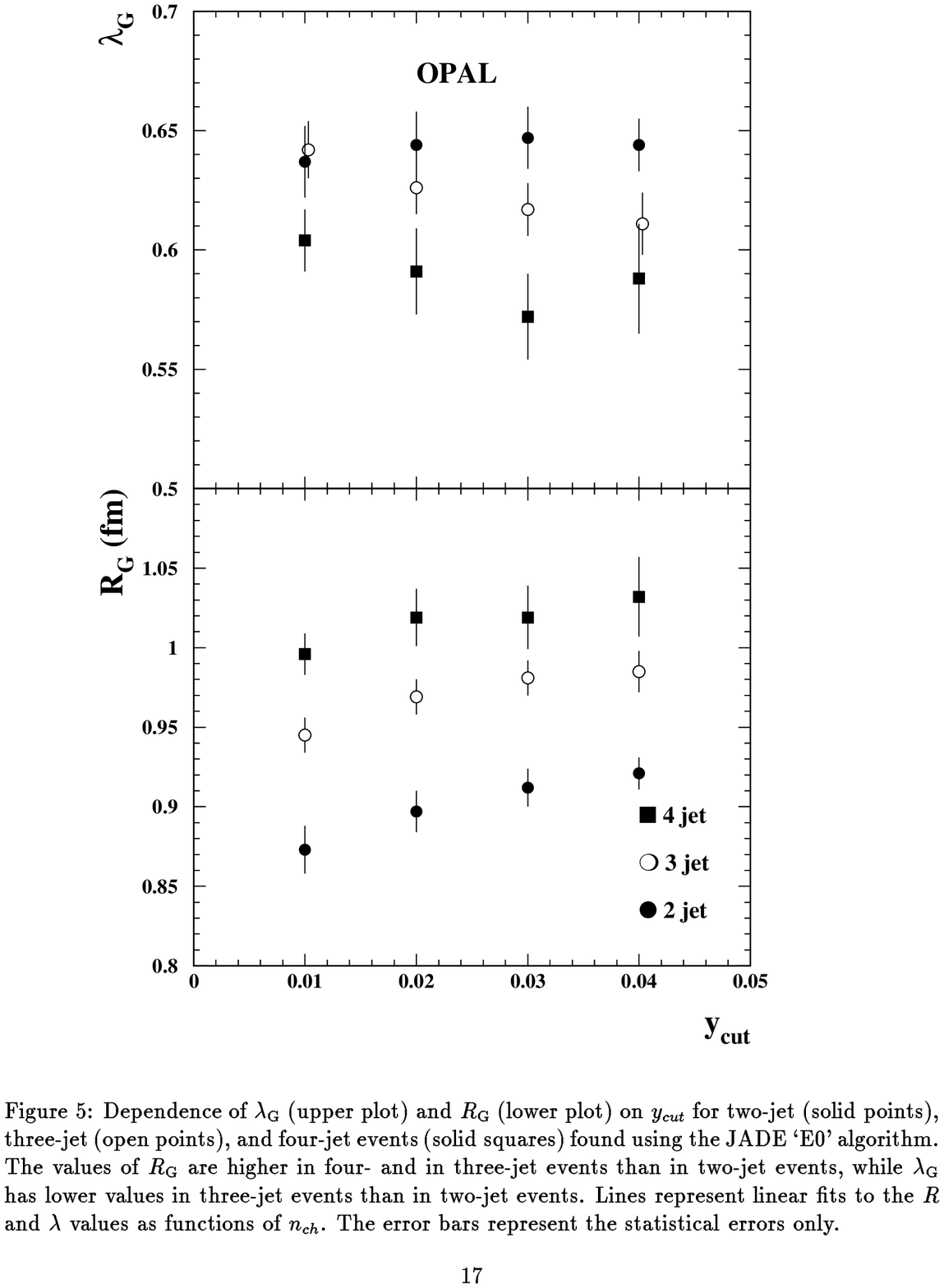}
  \hfil
   \includegraphics*[width=0.32\textwidth,bbllx=53,bblly=280,bburx=282,bbury=638]{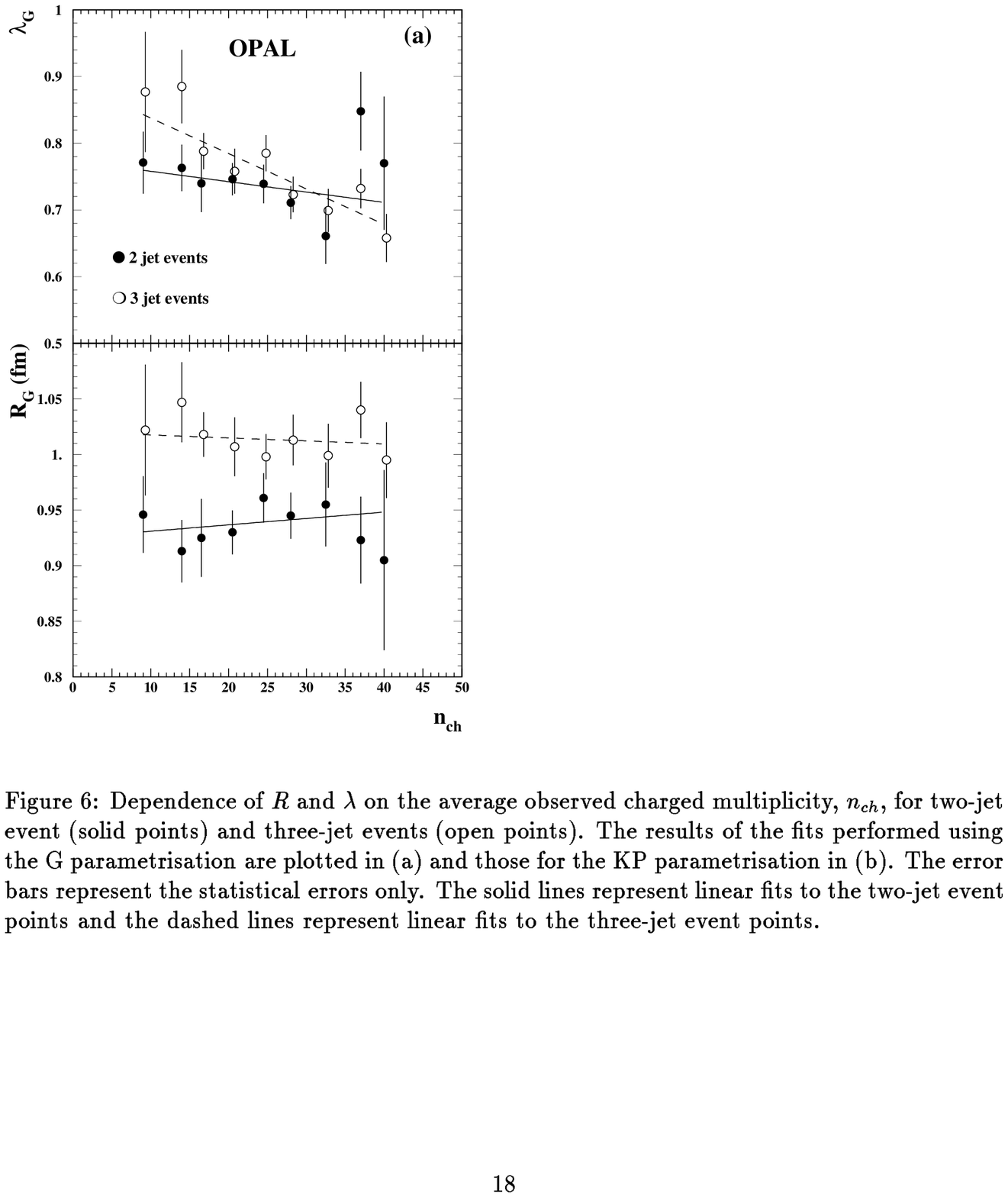}
  \caption{Dependence of $\lambda$ and $r$ from charged-pion 2-particle BEC
           on the charged particle multiplicity, $n_\mathrm{ch}$,
           and on the number of jets \cite{OPALmult:1996}.
          }
  \label{fig:mult}
\end{figure}

\paragraph{Dependence on the particle multiplicity}
The values of $\lambda$ and $r$ from charged-pion 2-particle BEC  depend on the charged particle
multiplicity, $n_\mathrm{ch}$,
of the events. As seen in Fig.~\ref{fig:mult}, $\lambda$ decreases with $n_\mathrm{ch}$ while $r$ increases.
However, a similar dependence is also seen on the number of jets.
When 2-jet events are selected, little, if any, dependence on $n_\mathrm{ch}$ remains.
For 3-jet events, $r$ seems independent of $n_\mathrm{ch}$, although $\lambda$ does decrease with
$n_\mathrm{ch}$.  Thus, the dependence of  $\lambda$ and $r$ on $n_\mathrm{ch}$ seems to be largely due to a
dependence on the number of jets.
 
\begin{figure}
\parbox{.4\textwidth}{
  \includegraphics*[width=0.33\textwidth,bbllx=100,bblly=46,bburx=440,bbury=435]{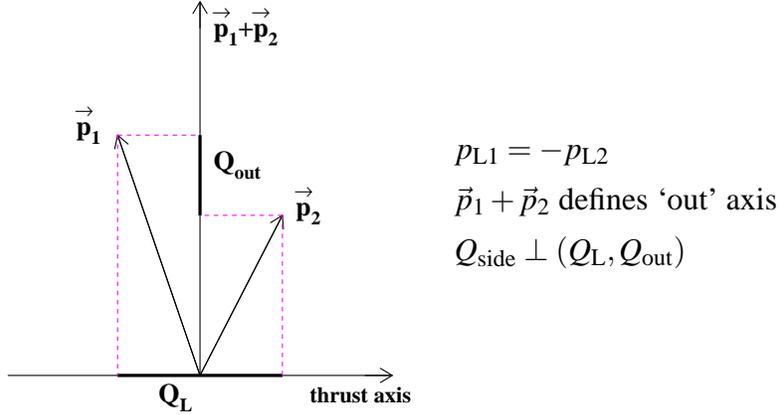}
 }
\parbox{.3\textwidth}{
{ $p_\mathrm{L1} = -p_\mathrm{L2}$}  \\[2mm]
$\vec{p}_1 + \vec{p}_2$ defines `out' axis \\[2mm]
{$\Qside\perp(\Qlong,\Qout)$}}
  \caption{The definition of the longitudinal center of mass system (LCMS).}
  \label{fig:lcms}
\end{figure}

\paragraph{Elongation of the source}
The Gaussian parametrization (\ref{Rtwo_G})  assumes a spherical source.  Given the jet
structure of \Pee\ events, one might expect a more ellipsoidal shape. To investigate this, the
parametrization is generalized to allow different radii for the direction along and perpendicular to the jet
axis.  The analysis is perfomed in the longitudinal center of mass system (LCMS), which is defined as
follows:  The pion pair is boosted along the jet-axis (taken, \eg, as the thrust axis), to a frame where the
sum of the longitudinal momenta of the two pions is zero.  The transverse axes, called ``out'' and ``side''
are defined such that the out direction is along the vector sum of the two momenta, $\vec{p_1}+\vec{p_2}$,
and the side direction completes the Cartesian coordinate frame.  This is illustrated in Fig.~\ref{fig:lcms}.
 
In the LCMS,
\begin{eqnarray}
           Q^2   &=&  \Qlong^2 + \Qside^2 + \Qout^2 - (\Delta E)^2  \nonumber \\
                 &=&  \Qlong^2 + \Qside^2 + \Qout^2 (1-\beta^2) , \quad\mathrm{where}\quad
                      \beta\equiv\frac{p_{\mathrm{out}\,1}+p_{\mathrm{out}\,2}}{E_{1}+E_{2}} \ .
                      \label{eq:qlcms}
\end{eqnarray}
The advantage of the LCMS is immediately apparent:  The energy difference, and therefore the difference in
emission time of the pions, couples only to the out component, \Qout.
Thus \Qlong\ and \Qside\ reflect only spatial dimensions of the source, while
\Qout\ reflects a mixture of spatial and temporal dimensions.
The Gaussian parametrization (\ref{eq:Rtwo_Glin}) becomes
\begin{equation}  \label{eq:Rtwo_Glin_LCMS}
    R_2(\Qlong,\Qside,\Qout)  = \gamma \left(1 + \lambda G\right)
                \left(1+\delta\Qlong+\epsilon\Qout+\xi\Qside\right) \ .
\end{equation}
Assuming an azimuthally symmetric Gaussian
shape, there is only one non-zero off-diagonal term, and $G$ is given by
\begin{equation}  \label{eq:G_LCMS}
         G = \exp \left(-r^2_\mathrm{L}{\Qlong^2}
                        -r^2_\mathrm{out}{\Qout^2}
                        -r^2_\mathrm{side}{\Qside^2}
                        +2\rho_\mathrm{L,out}{\Rlong\Rout}{\Qlong\Qout}
                     \right)      \ .
\end{equation}
Such an analysis has been performed by \Lthree\ \cite{L3_3D:1999} and \OPAL\ \cite{OPAL3D:2000}.
In fact, $\rho_\mathrm{L,out}$ turns out consistent with zero, and is then fixed to zero in the fits.
Two-dimensional LCMS analyses have been performed by \ALEPH\ \cite{ALEPH:2004} and \DELPHI\ \cite{DELPHI2D:2000},
in which the out and side components are replaced by a transverse one, $Q^2_\mathrm{t}=\Qout^2+\Qside^2$.
However, the interpretation of the corresponding parameter, \Rtrans, as a transverse radius is not
unambiguous, since it includes the effect of the difference in time of emission.
Both \Lthree\ and \ALEPH\ fit not only (\ref{eq:Rtwo_Glin_LCMS}), but a similar expression where $G$ is
replaced by a lowest-order Edgeworth expansion, which gives a better fit (\eg, in \Lthree, a confidence
level of 30\% compared with 3\% for the Gaussian fit).
The ratio of transverse to longitudinal radii found in the four experiments are shown in
Table~\ref{tab:elong}.
The longitudal radius is clearly about 20\% larger than the transverse radius.

\begin{table}
\begin{tabular}{ccccr@{$\pm$}c@{$\pm$}lr@{{$\pm$}}c@{{$\pm$}}l}
  \hline
 Experiment& Data   &Reference&  Gauss / & \multicolumn{3}{c}{2-D} & \multicolumn{3}{c}{3-D} \\
           &         &sample & Edgeworth & \multicolumn{3}{c}{$\Rtrans/\Rlong$} &
                                                  \multicolumn{3}{c}{$\Rside/\Rlong$} \\
  \hline
   \DELPHI & 2-jet   & mixed & Gauss  & 0.62 & 0.02 & 0.05 &  \multicolumn{3}{c}{---} \\
  \hline
   \ALEPH  & 2-jet   & mixed & Gauss  & 0.61 & 0.01 & 0.?? &  \multicolumn{3}{c}{---} \\
           & 2-jet   & +\,-- & Gauss  & 0.91 & 0.02 & 0.?? &  \multicolumn{3}{c}{---} \\
           & 2-jet   & mixed &Edgeworth&0.68 & 0.01 & 0.?? &  \multicolumn{3}{c}{---} \\
           & 2-jet   & +\,-- &Edgeworth&0.84 & 0.02 & 0.?? &  \multicolumn{3}{c}{---} \\
  \hline
   \OPAL   & 2-jet   & +\,-- & Gauss  &\multicolumn{3}{c}{---}&{0.82}&{0.02}&${^{0.01}_{0.05}}$\\
  \hline
   \Lthree &  all    & mixed & Gauss  &\multicolumn{3}{c}{---}&{0.80}&{0.02}&${^{0.03}_{0.18}}$\\
           &  all    & mixed&Edgeworth&\multicolumn{3}{c}{---}&{0.81}&{0.02}&${^{0.03}_{0.19}}$\\
  \hline
\end{tabular}
\caption{Ratio of transverse or side radius to longitudinal radius from charged-pion 2-particle BEC analyses
  of retail organisation}
\label{tab:elong}
\end{table}
 
\begin{figure}
  \includegraphics*[width=.5\textwidth]{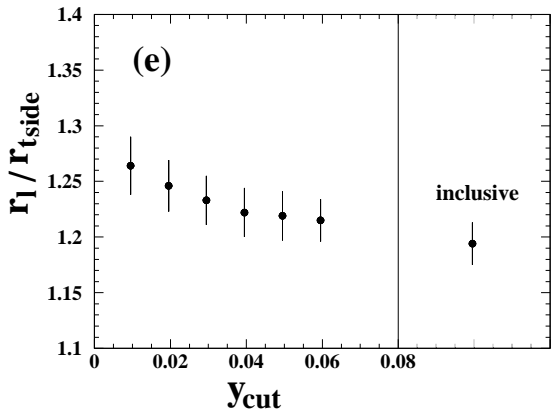}
  \caption{The ratio of \Rlong\ to \Rside\ for 2-jet events as function of the jet resolution parameter, \ycut.}
  \label{fig:elongycut}
\end{figure}

In Fig.~\ref{fig:elongycut} we see that the amount of this elongation increases when
narrower 2-jet events are selected.
 
It should also be mentioned that \ZEUS\ \cite{ZEUS2D:2004} performed a similar 2-dimensional analysis
in deep inelastic ep interactions.  The ratio $\Rtrans/\Rlong$ found, similar to that found by \DELPHI\ and
\ALEPH, is independent of the virtuality of the exchanged photon.
 
\paragraph{\Pgpz\Pgpz}
In hadronization models with local charge conservation, \eg, string models, neutral pions can be produced
closer together than identical charged pions.  One could expect this to be reflected in a smaller BEC radius
for \Pgpz\Pgpz\ than for $\pi^\pm\pi^\pm$.  Only two \Pee\ experiments have attempted a BEC analysis of
\Pgpz\Pgpz, \Lthree\ \cite{L3_pi0:2002} and \OPAL\ \cite{OPALpi0:2003}.
The experimental selections used in the two analyses are quite different, dictated as they are by the
characteristics of the different detectors.  While \Lthree\ requires the energy of the pions to be less than
6\,\GeV, \OPAL\ demands the \Pgpz\ momenta to be greater than 1\,\GeV.  Further, \OPAL\ uses only 2-jet
events, defined as having thrust larger than 0.9.
 
\begin{figure}
  \includegraphics[width=.48\textwidth]{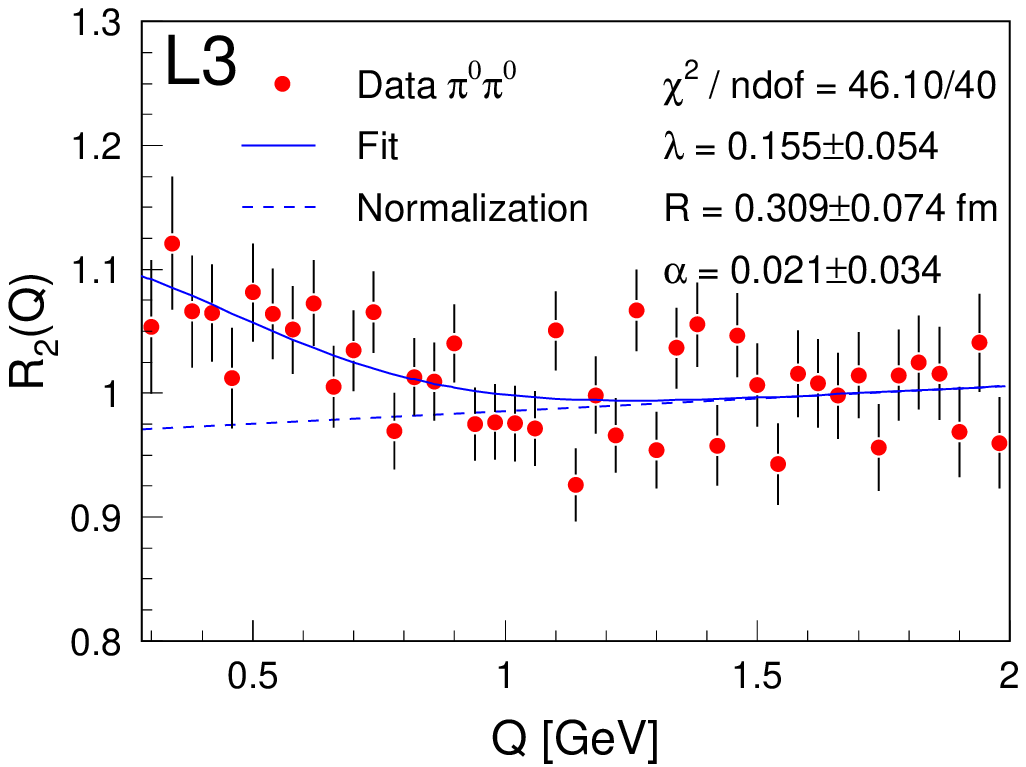}
  \includegraphics[width=.48\textwidth]{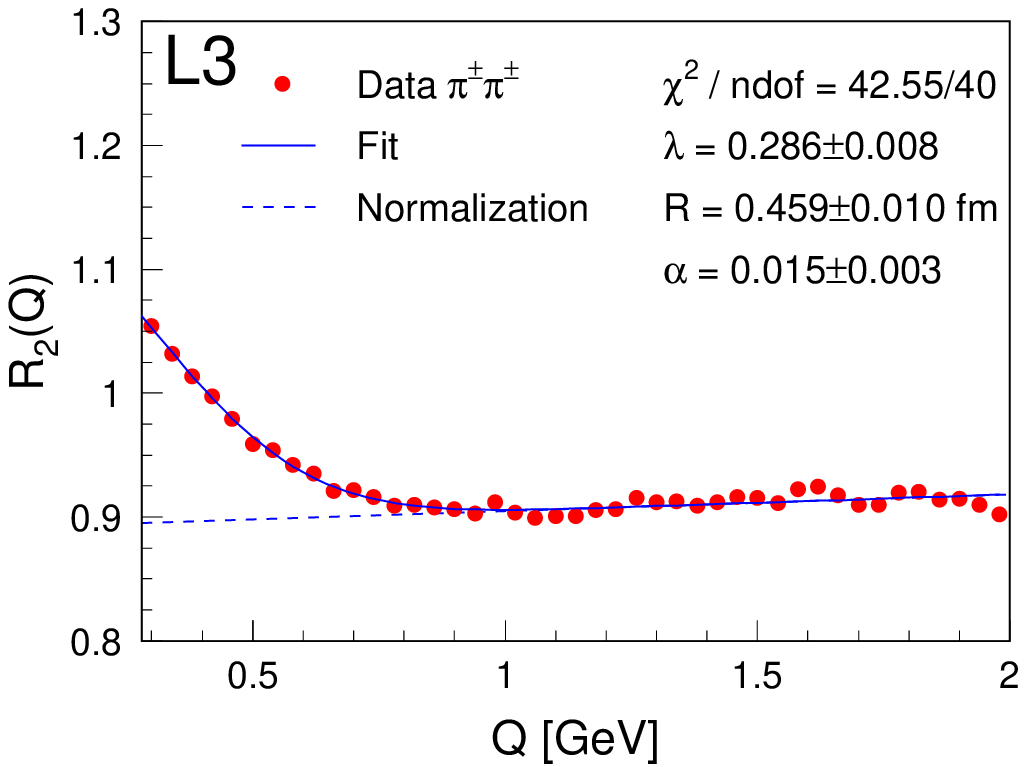}
  \caption{$R_2$ distribution for \Pgpz\Pgpz\ and for $\pi^\pm\pi^\pm$ \cite{L3_pi0:2002}.}
  \label{fig:L3pi0}
\end{figure}

\begin{figure}
  \includegraphics[width=.55\textwidth,bblly=20,bburx=519,bbury=519]{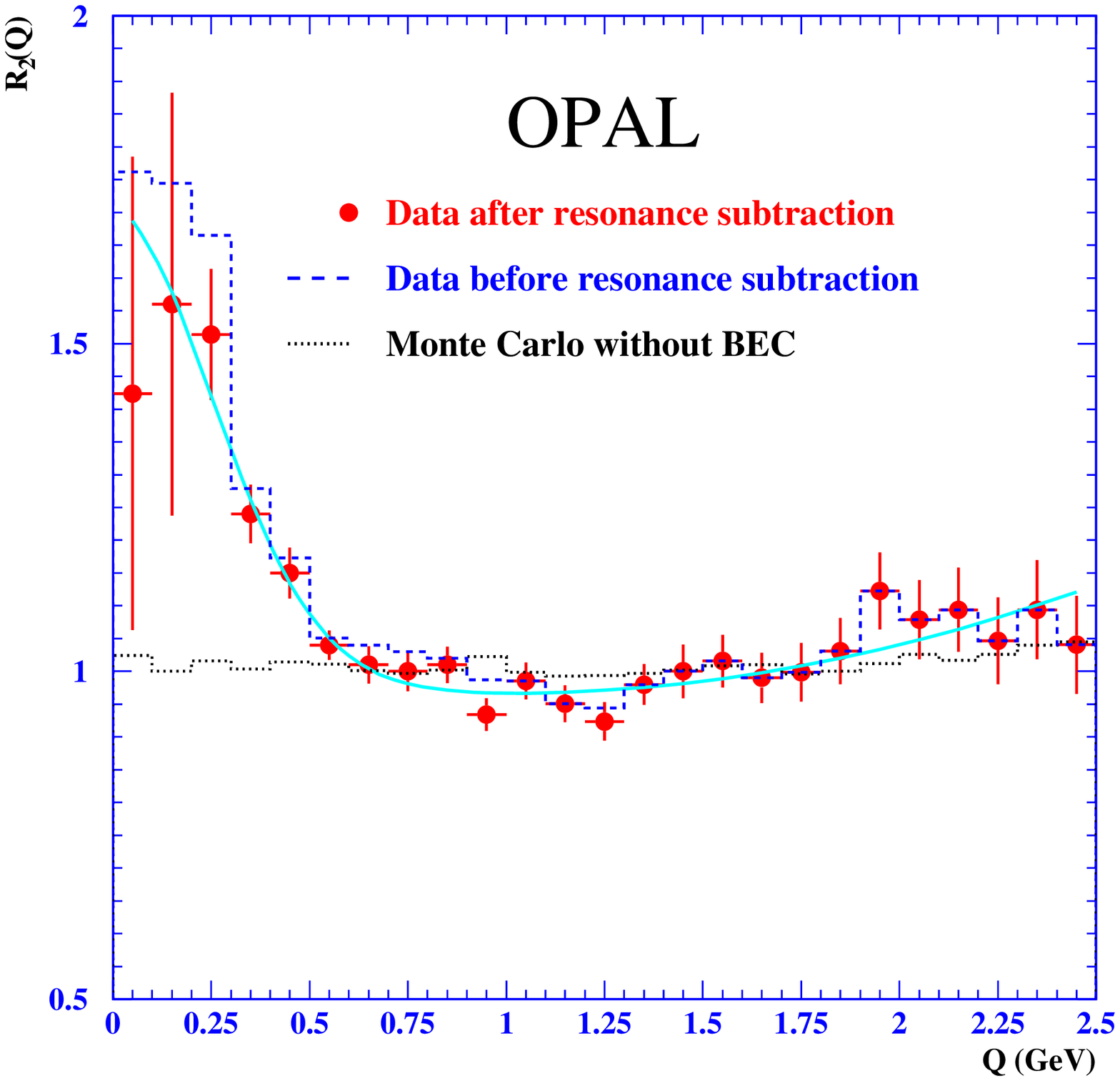}
  \caption{$R_2$ distribution for \Pgpz\Pgpz\ \cite{OPALpi0:2003}.}
  \label{fig:OPALpi0}
\end{figure}

The $R_2$ distributions and fits are shown in Figs.~\ref{fig:L3pi0} and~\ref{fig:OPALpi0}, respectively.
In order to make a comparison with charged pions, \Lthree\ also analyzes $\pi^\pm\pi^\pm$ with a
selection similar to its \Pgpz\ selection.
The resulting   $R_2$ is also shown in Fig.~\ref{fig:L3pi0}.
The results of fits to these distributions, as well as two other $\pi^\pm\pi^\pm$ results are listed
in Table~\ref{tab:pi0}.
 
Comparison of the \Lthree\ values of $r$ for \Pgpz\Pgpz\ and $\pi^\pm\pi^\pm$ with the same selection
indicates that the radius is smaller in the \Pgpz\Pgpz\ case, but the significance is only about 1.5
standard deviations. $\lambda$ is also smaller in the \Pgpz\Pgpz\ case, with similar significance.
 
For \OPAL, the comparison is more difficult to make, since \OPAL's charged-pi results use a different
reference sample and different selection.
Other experiments have found that the ratio of $r$ using a mixed reference sample to that using unlike-sign
is about 0.68 (\ALEPH\ \cite{ALEPH:2004}) or 0.56 (\DELPHI\ \cite{DELPHI:1992}).  Applying such a factor would
lower the \OPAL\ value to about 0.62, which agrees well with the \OPAL\ result for \Pgpz\Pgpz.
However, it is not clear what the effect of the 2-jet and \Pgpz-momentum cuts is.  In the \Lthree\ case,
requiring the pions to have $E<6\,\GeV$ and using a Monte Carlo reference sample rather than a
mixed one decreased $r$ by about 30\%.  It is therefore conceivable that the \OPAL\ requirement of
$p>1\,\GeV$ would increase $r$, in which case $r$ for \Pgpz\Pgpz\ would be smaller than for
$\pi^\pm\pi^\pm$.

\begin{table}
  \begin{tabular}{lllccc}
    \hline
          &             &                     & Reference &          &      \\
          & Experiment  & Selection           & sample    & {$r$} (fm)         & $\lambda$ \\
\hline
 $\Pgpz\Pgpz$ & \Lthree\ & $E_\Pgp<6\,\GeV$          & MC & $0.31\pm0.10$      & $0.16\pm0.09$ \\
            &\OPAL\      & $p_\Pgp>1\,\GeV$, 2-jet  & mix & $0.59\pm0.09$      & $0.55\pm0.14$ \\ \hline
 $\pi^\pm\pi^\pm$
          & \Lthree\     & $E_\Pgp<6\,\GeV$         & MC  & $0.46\pm0.01$          & $0.29\pm0.03$ \\
          & \Lthree      &                          & mix & $0.65\pm0.04$          & $0.45\pm0.07$ \\
          & \OPAL        &                          & +-- & $1.00^{+0.03}_{-0.10}$ & $0.57\pm0.05$ \\
          \hline
  \end{tabular}
 
\caption{Results of fits to $R_2$ for \Pgpz\Pgpz\ \cite{L3_pi0:2002,OPALpi0:2003}
         and $\pi^\pm\pi^\pm$ \cite{L3_pi0:2002,L3_3pi:2002,OPAL3D:2000}.
         The indicated uncertainties combine statistcal and systematic uncertainties.
        }
\label{tab:pi0}
\end{table}

\subsection{3-particle BEC}
 
There have also been analyses of BEC among three charged pions.
As mentioned in the Introduction, with three (or more) particles the correlations may be classified as
trivial, simply a consequence of lower-order correlations, and genuine.  We study these correlations using
$R_3(Q_3)=\rho(Q_3)/\rho_0(Q_3)$.  Note that $Q_3^2=Q^2_{12}+Q^2_{23}+Q^2_{31}$.
The same assumptions that lead to (\ref{Rtwo_G}) for $R_2$, lead to \cite{Lorstad:1989,Lyuboshitz:1991}
\begin{eqnarray}
  R_2(Q_{ij})               &=& 1 + \lambda\abs{G(q_{ij})}   \label{eq:Rtwo_Gij}  \\
  R_3(Q_{12},Q_{23},Q_{31}) &=& 1 +
           {\underbrace{\lambda
             \left(|{G(Q_{12})}|^2
                 + |{G(Q_{23})}|^2
                 + |{G(Q_{31})}|^2\right)}_\mathrm{from\ 2\mbox{-}particle\ BEC}} \nonumber \\
         \hspace{-4mm} &\hspace{-2mm} & 
         \phantom{1}
                    + {
\underbrace{2{\lambda^{1.5}}\,\Re\{{G(Q_{12})}
                                   {G(Q_{23})}
                                   {G(Q_{31})}\}}_\mathrm{from\
                        genuine\ 3\mbox{-}particle\ BEC} }   \ ,         \label{eq:Rthree_G}
\end{eqnarray}
where $G(Q)\!=\!\int\!\dd{x}\,e^{\imath Qx} S(x)=\abs{G}e^{\imath\phi}$ is the Fourier transform of the
source density, and $\Re$ denotes the real part.
Note that $R_3$, unlike $R_2$, depends on the phase of $G(Q)$.
We define $R_3^{\mathrm{gen}}$ as the $R_3$ that would occur if there were no 2-particle BEC:
\begin{equation}
  R_3^{\mathrm{gen}}(Q_{12},Q_{23},Q_{31}) =  1 +
            2{\lambda^{1.5}}\,\Re\{{G(Q_{12})}
                                   {G(Q_{23})}
                                   {G(Q_{31})}\}       \ .                \label{eq:Rthree_gen}
\end{equation}
Defining
\begin{equation}  \label{eq:omega_def}
    \omega={\cos(\phi_{12}+\phi_{23}+\phi_{13})} \ ,
\end{equation}
the above equations yield
\begin{equation}  \label{eq:omega}
    \omega = \frac{R_3^\mathrm{gen}(Q_3) - 1}
                  {2\sqrt{(R_2(Q_{12})-1)(R_2(Q_{23})-1)(R_2(Q_{13})-1)}} \ ,
\end{equation}
which simplifies in the case of a Gaussian to
\begin{equation}  \label{eq:omega_G}
    \omega = \frac{R_3^\mathrm{gen}(Q_3) - 1}
                  {2\sqrt{R_2(Q_3)-1}}  \ .
\end{equation}
If the particle production is completely incoherent, the phase $\phi_{ij}$ is expected to be zero, and
consequently we should find $\omega=1$.  As we have seen, incoherence also implies $\lambda=1$, but
$\lambda$ is affected by many other factors, which should not affect $\omega$.
 
\begin{figure}
  \includegraphics*[width=.9\textwidth,bbllx=34,bblly=51,bburx=672,bbury=388]{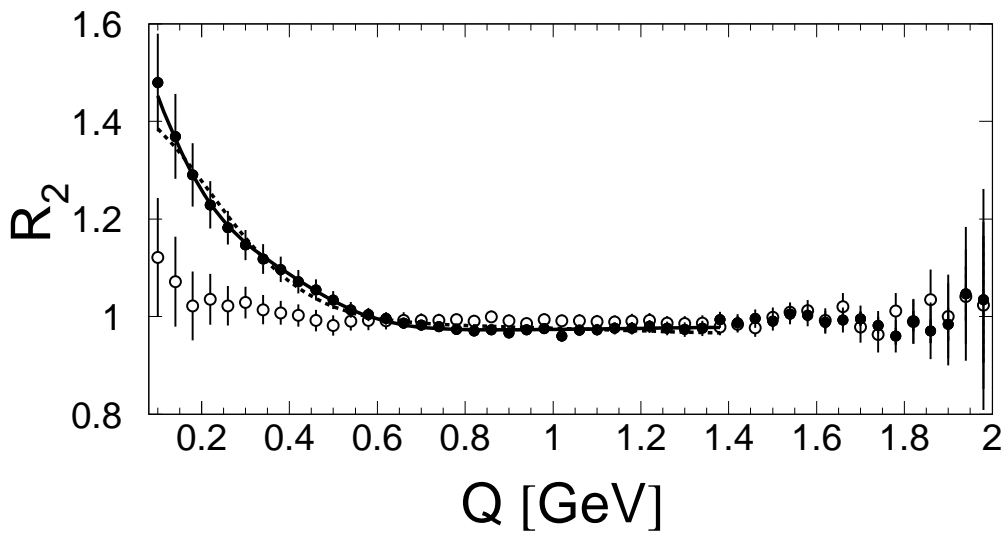}
  \caption{$R_2(Q)$ measured by \Lthree\ \cite{L3_3pi:2002}. The full circles represent the data, while the
           open circles correspond to results from a Monte Carlo model without BEC.  The dashed and full
           lines show the fits of a Gaussian and an Edgeworth expansion, respectively.}
  \label{fig:3piR2}
\end{figure}

\begin{figure}
  \includegraphics*[width=.9\textwidth,bbllx=27,bblly=58,bburx=787,bbury=938]{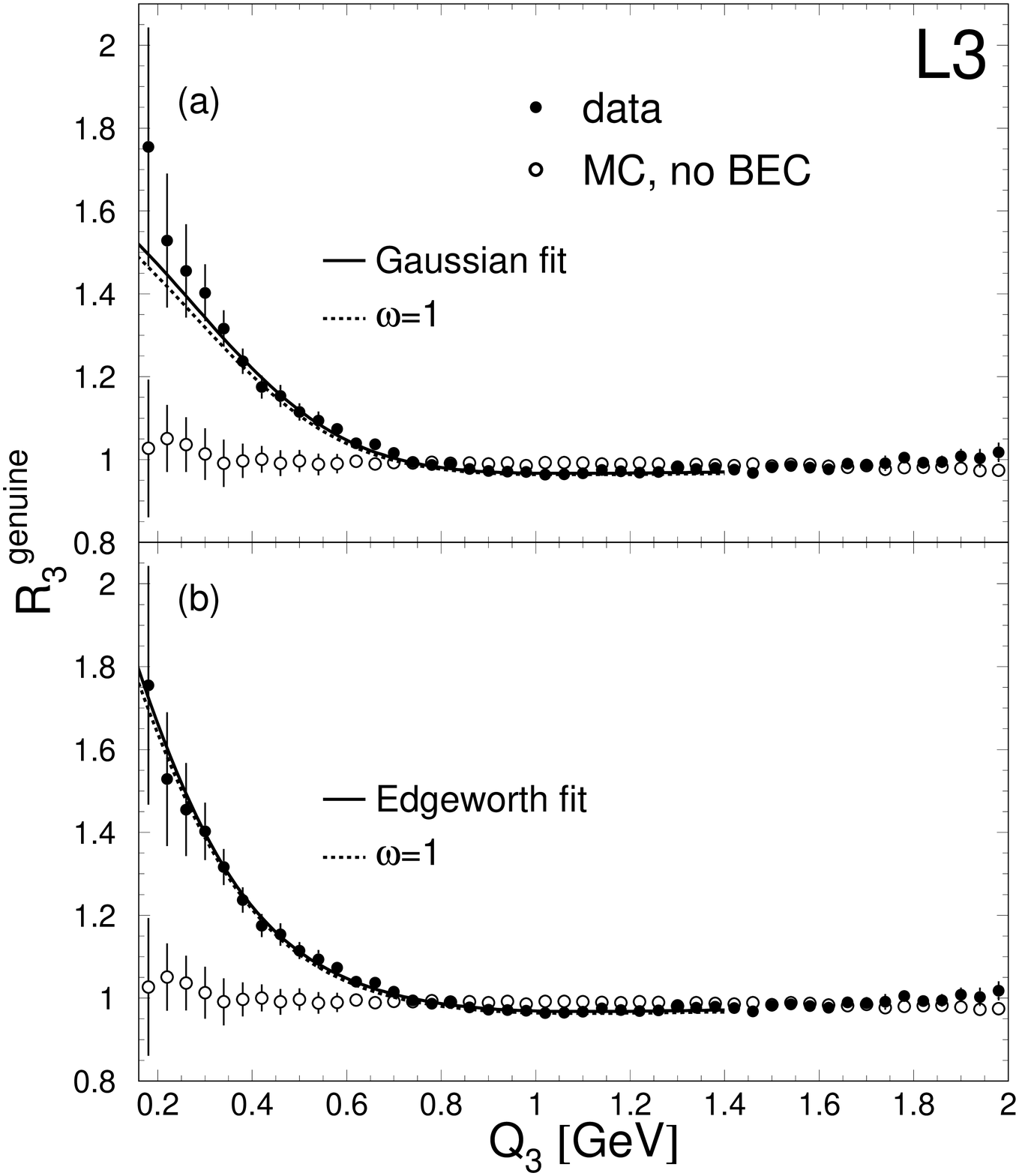}
  \caption{$R_3^{\mathrm{gen}}(Q_3)$ measured by \Lthree\ \cite{L3_3pi:2002}.
           The full circles represent the data, while the
           open circles correspond to results from a Monte Carlo model without BEC.
           The full line in (a) is a fit of a Gaussian parametrization, while
                      in (b) it is a fit of an Edgeworth parametrization.
           The dashed lines show the expectation from (\ref{eq:Rthree_gen}) with the results of the fits to
           $R_2$ assuming $\omega=1$.}
  \label{fig:3piR3}
\end{figure}

The \Lthree\ measurements \cite{L3_3pi:2002} of $R_2$ and $R_3^\mathrm{gen}$ in hadronic \PZ\ decays
are shown in
Figs.~\ref{fig:3piR2} and~\ref{fig:3piR3}. The distributions are fit with both a Gaussian and an
Edgeworth parametrization.  The Edgeworth parametrization provides a better fit of the data.
It is reassuring to note that the values of $\lambda$ and $r$ obtained from the fits to $R_2$ and
$R_3^\mathrm{gen}$, which are listed in Table~\ref{tab:3pi}, agree perfectly.
 
\begin{table}
\begin{tabular}{lccc}
  \hline
     fit to           &           & Gaussian             & Edgeworth           \\
  \hline
           $R_2$      & $\lambda$ & $0.45\pm0.06\pm0.03$ & $0.72\pm0.08\pm0.03$ \\
   $R_3^\mathrm{gen}$ &           & $0.47\pm0.07\pm0.03$ & $0.75\pm0.10\pm0.03$ \\
  \hline
   $R_2$              & $r$       & $0.65\pm0.03\pm0.03$ & $0.74\pm0.06\pm0.02$ \\
   $R_3^\mathrm{gen}$ & (fm)      & $0.65\pm0.06\pm0.03$ & $0.72\pm0.08\pm0.03$ \\
  \hline
\end{tabular}
 \caption{Results of fits to $R_2$ and $R_3^\mathrm{gen}$
          using Gaussian and Edgeworth parametrizations\ \cite{L3_3pi:2002}.}
\label{tab:3pi}
\end{table}

The dashed lines in Fig.~\ref{fig:3piR3} are the predictions of $R_3^\mathrm{gen}$ using
(\ref{eq:Rthree_gen}) with the results of the fits to $R_2$ assuming $\omega=1$.
These dashed lines are quite close to the solid lines, which represent fits to $R_3^{\mathrm{gen}}(Q_3)$,
confirming that $\omega=1$ is a reasonable hypothesis.
 
\begin{figure}
  \includegraphics[width=.70\textwidth]{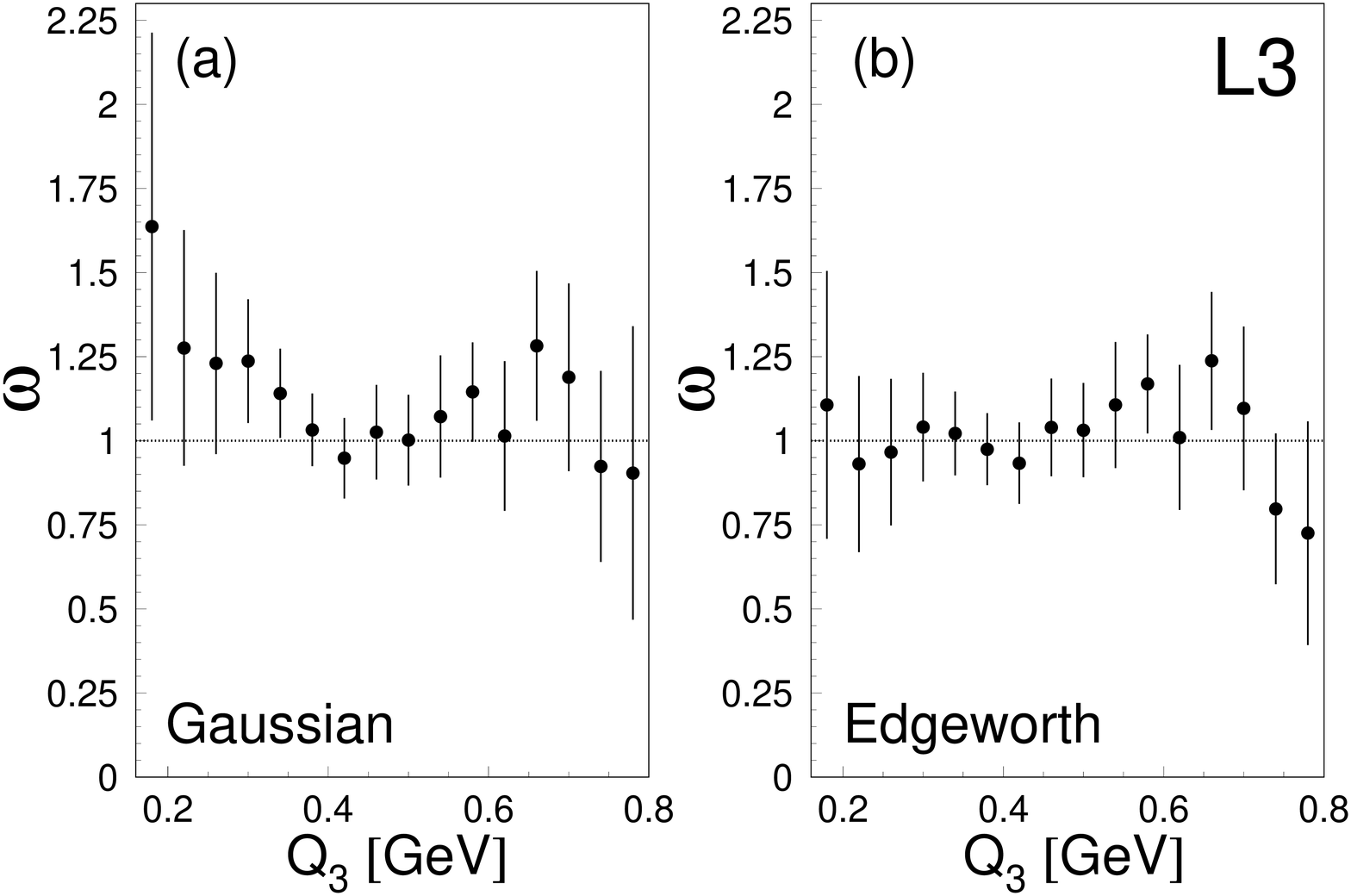}
  \caption{The ratio $\omega$ (\ref{eq:omega}) as function of $Q_3$
           computed using the measured $R_3^{\mathrm{gen}}(Q_3)$ and $R_2$ from its fits.
           }
  \label{fig:omega}
\end{figure}

Another way to examine this hypothesis is to compute $\omega$ using (\ref{eq:omega}) for each bin in $Q_3$
using the measured $R_3^{\mathrm{gen}}(Q_3)$ and $R_2$ from its fits.  The results are shown in
Fig.~\ref{fig:omega}.  We conclude that $\omega$ is perfectly consistent with unity, and consequently that
particle production is completely incoherent.
 
BEC among three pions have previously been observed at lower energies \cite{TASSO:1986,MarkII:1989}.
At \LEP, \DELPHI\ \cite{DELPHI3pi:1995} and \OPAL\ \cite{OPAL3pi:1998} have also observed genuine 3-pion BEC.
With the exception of \DELPHI, which did not do a comparable 2-pi analysis, they all report that the values
of $\lambda$ and $r$ obtained for 2- and 3-pions are consistent.  Unfortunately, none of these experiments
performed an analysis using $\omega$.

\subsection{BEC in hadronic W decays}
\begin{figure}
  \includegraphics[width=.5\textwidth]{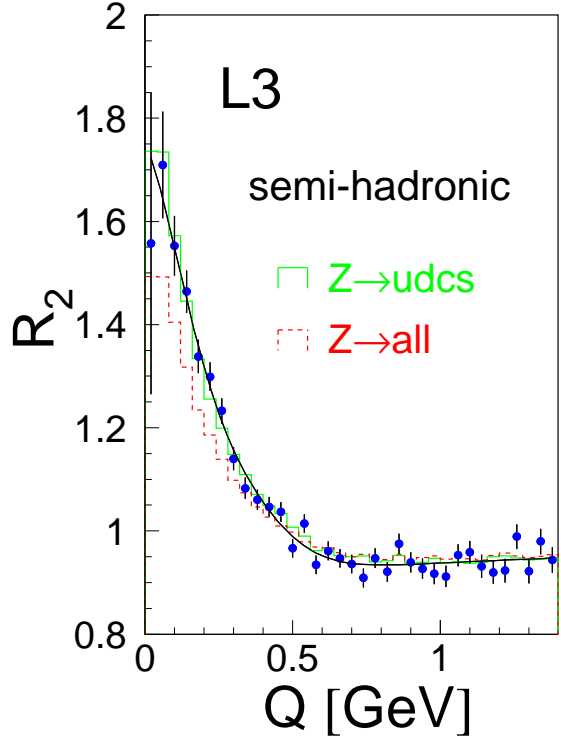}                \\
  \caption{$R_2(Q)$ measured by \Lthree\ \cite{L3_WW:2002} for
           hadronic W-boson decays from $\PeeWW\rightarrow\Pq\Paq\ell\nu$
           and hadronic \PZ-boson decays to all flavors and to udsc flavors only.
          }
  \label{fig:WZbec}
\end{figure}

\paragraph{$\PW\rightarrow\Pq\Paq$}
Having found no evidence for a center-of-mass energy dependence of BEC,
and noting that in any case the masses of the Z and W are not much different,
there is only one reason to expect BEC to be different in W decays
than in Z decays, namely the different flavor composition.  Whereas about 20\%
of Z decays are to \Pqb\Paqb, almost no W decays involve a \Pqb-quark.
The long lifetime of the \Pqb\ results in  diminished BEC.
Thus, we should expect BEC in W-decay to be like that in the decay of
the Z to light (udsc) quarks.  This is indeed found to be the case
\cite{ALEPH_WW:2000,L3_WW:2002}, as is shown in Fig.~\ref{fig:WZbec},
where $R_2(Q)$ for hadronic decays of the \PW-boson produced in
$\PeeWW\rightarrow\Pq\Paq\ell\nu$ is compared to that of \PZ-boson decays to all flavors
and to udsc flavors only.

\paragraph{Inter-string BEC}
A more interesting case is $\PeeWW\rightarrow\Pq\Paq\Pq\Paq$.
The \PW-bosons are produced not far above threshold and consequently travel only about 0.7\,fm before
decaying. This is smaller than the distance over which hadronization occurs.  Therefore one expects a
significant degree of overlap of the two hadronizing systems, resulting in BEC not only between particles
produced by the same \PW, but also between particles produced by different \PW-bosons.
On the other hand, in the string picture no BEC is expected between particles arising from different
strings.
 
Whether or not this so-called inter-string BEC exists is thus a fundamental question for the string picture.
It is also important for the measurement of properties of the \PW-boson, in particular its mass \cite{LS95}.
Improper simulation of inter-string BEC in Monte Carlo programs would lead to a bias in the mass measurement.
 
All four \LEP\ experiments have studied this question
\cite{ALEPH_WW:2000,ALEPH_WW:2005,DELPHI_WW:1997,DELPHI_WW:2005,L3_WW:2002,OPAL_WW:2004}.
The basic method \cite{CWK99} is to test the expectation of no inter-string BEC.
If the two \PW-bosons decay completely independently, the two-particle density in the 4-quark channel
is given by
\begin{eqnarray*}
   {\rho_\mathrm{4q}(p_1,p_2)} = & {\rho^{+}(p_1,p_2)}          & \mathrm{1,2\ from\ \PWp}  \\
                               + & {\rho^{-}(p_1,p_2)}          & \mathrm{1,2\ from\ \PWm}  \\
                               + & {\rho^{+}(p_1)\rho^{-}(p_2)} & \mathrm{1\ from\ \PWp,\ 2\ from\ \PWm} \\
                               + & {\rho^{+}(p_2)\rho^{-}(p_1)} & \mathrm{1\ from\ \PWm,\ 2\ from\ \PWp.}
\end{eqnarray*}
Assuming {$\rho^+ = \rho^- = \rho_\mathrm{2q}$}, which would only be strictly true if there is complete
overlap,
\begin{equation}  \label{eq:WW1}
   \rho_\mathrm{4q}(p_1,p_2) =  2\rho_\mathrm{2q}(p_1,p_2)
                             +  2\rho_\mathrm{2q}(p_1)\rho_\mathrm{2q}(p_2)  \ .
\end{equation}
The density $\rho_\mathrm{4q}(p_1,p_2)$ is measured in $\PeeWW\rightarrow\Pq\Paq\Pq\Paq$ and
            $\rho_\mathrm{2q}(p_1,p_2)$             in $\PeeWW\rightarrow\Pq\Paq\ell\nu$.
The remaining term,  $\rho_\mathrm{2q}(p_1)\rho_\mathrm{2q}(p_2)$, is estimated by
$\rho_\mathrm{mix}(p_1,p_2)$
obtained  by  mixing  $\ell^+\nu\Pq\Paq$ and $\Pq\Paq\ell^-\nu$  events
after removal of the $\ell^+$ and~$\ell^-$.
Thus, (\ref{eq:WW1}) becomes
\begin{equation}  \label{eq:WW}
           \rho_\mathrm{4q}(Q) = 2\rho_\mathrm{2q}(Q) + 2\rho_\mathrm{mix}(Q)  \ .
\end{equation}
Various quantities are defined to test the validity of (\ref{eq:WW}):
\begin{eqnarray}   \label{eq:WWrho}
         \Delta\rho(Q) &=& \rho_\mathrm{4q}(Q)
                          -  \left[ 2{\rho_\mathrm{2q}(p_1,p_2)}
                                  + 2{\rho_\mathrm{mix}(p_1,p_2)} \right]
    \\             \label{eq:WWD}
                  D(Q) &=& \frac{\rho_\mathrm{4q}(Q)}
                                {2\rho_\mathrm{2q}(Q) + 2\rho_\mathrm{mix}(Q)}
    \\             \label{eq:WWdI}
            \deltaI(Q) &=& \frac{\Delta\rho(Q)}
                                {2\rho_\mathrm{mix}(Q)}   \ .
\end{eqnarray}
The quantity $\deltaI(Q)$ is actually the correlation function of genuine inter-W BEC \cite{DeWolf:2001}
and is thus not only a test of no inter-W BEC, but the quantity of interest if inter-W
BEC indeed exists.
 
\begin{figure}
    \includegraphics[width=.48\textwidth]{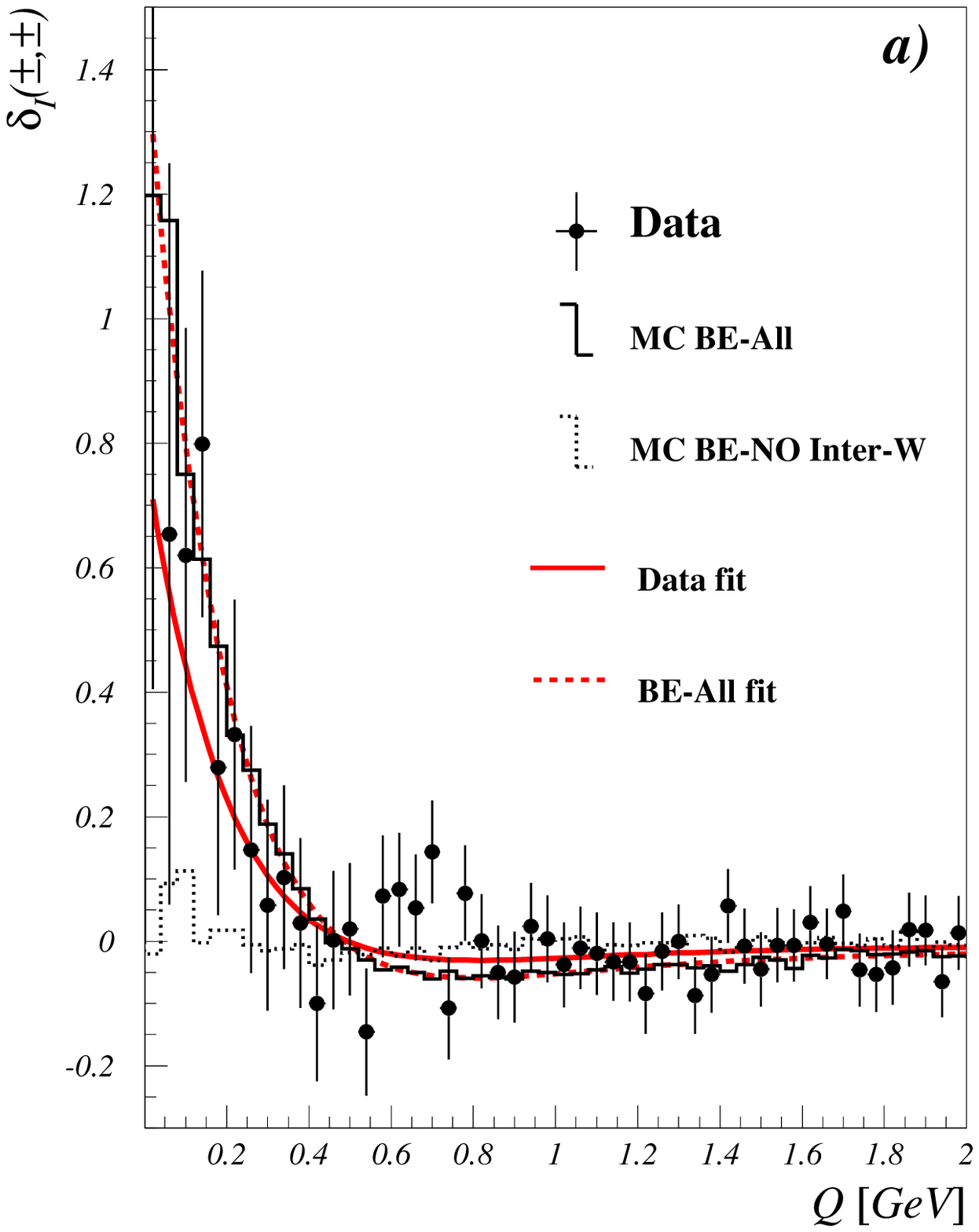}
    \includegraphics[width=.48\textwidth]{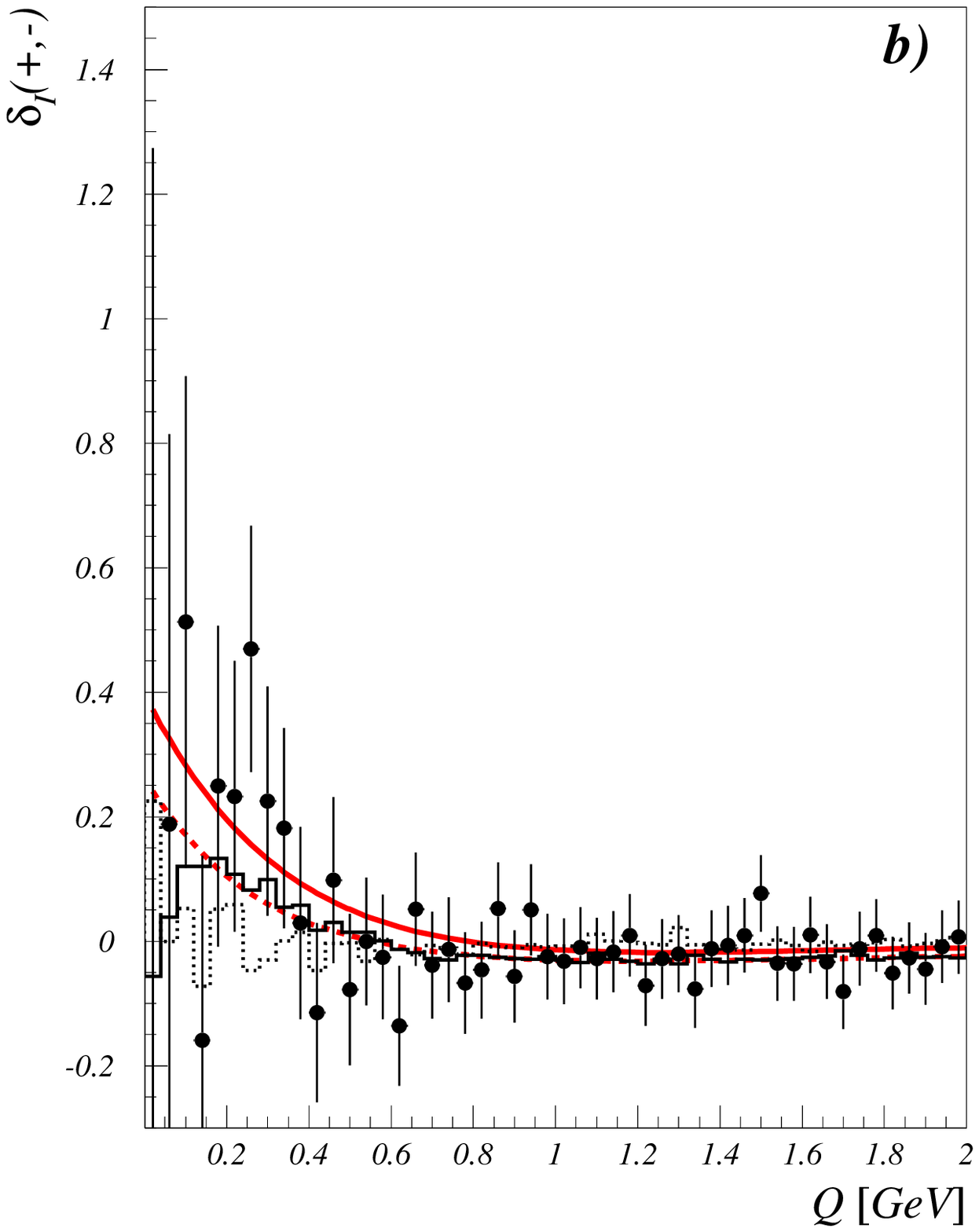}
  \caption{$\deltaI(Q)$ measured by \DELPHI\ \cite{DELPHI_WW:2005} for
           (a) like- and (b) unlike-sign pion pairs.
           Also shown are the expectations of Monte Carlo models including BEC among all pions or including
           BEC only among pions from the same W.
          }
  \label{fig:DELPHI_WW}
\end{figure}

All four \LEP\ experiments have now reported final results
\cite{ALEPH_WW:2005,DELPHI_WW:2005,L3_WW:2002,OPAL_WW:2004}.
Fig.~\ref{fig:DELPHI_WW} shows results of \DELPHI\ \cite{DELPHI_WW:2005}, the only experiment which claims
to have seen significant inter-W BEC.
The \deltaI\ distribution of like-sign pions (Fig.~\ref{fig:DELPHI_WW}a) clearly shows an
enhancement at small $Q$, while no enhancement is seen in the Monte Carlo
distribution where BEC, as modeled by the \BEtt\ model \cite{LS98}, is included only between pions from the
same~\PW.  When pions from different \PW-bosons are modeled with the same BEC as pions from the same \PW, a
larger enhancement is seen than that of the data.
 
However, the interpretation of the enhancement in the data as inter-W BEC is somewhat clouded by the
observance of an enhancement, albeit smaller, in the \deltaI\ distribution of unlike-sign pions
(Fig.~\ref{fig:DELPHI_WW}b).  In this distribution the enhancement in the data is larger than that for Monte
Carlo with inter-W BEC.
 
\begin{figure}
  \includegraphics[width=.6\textwidth]{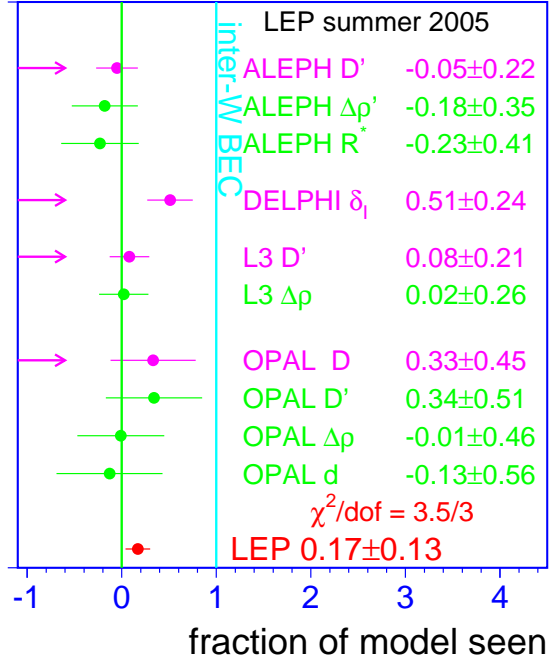}
  \caption{The amount of inter-W BEC observed expressed as a fraction of the prediction of the \BEtt\ model.
           The measurements indicated by arrows are combined to give the preliminary \LEP\ result
           \cite{LEPWWBEC:2005}.
          }
  \label{fig:WWcomp}
\end{figure}

The final results for the four \LEP\ experiments are compared in Fig.~\ref{fig:WWcomp}.
Here the results are expressed as the ratio of the effect seen to that expected in the \BEtt\ model
with the same inter-W BEC as intra-W BEC.
The measurements indicated by arrows are combined to give the preliminary \LEP\ result
\cite{LEPWWBEC:2005} of $0.17\pm0.13$.  The data are thus compatible with no inter-W BEC.
 
\subsection{BEC in the Lund string model}
\begin{figure}
  \includegraphics*[width=.6\textwidth,bbllx=172,bblly=308,bburx=434,bbury=490]{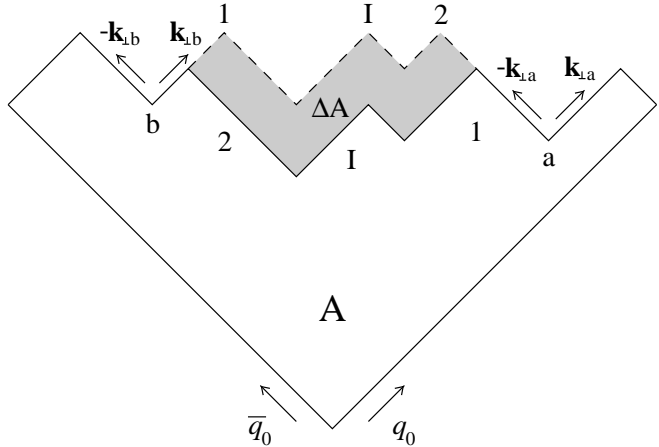}
  \caption{The break-up of a Lund string into hadrons, showing two break-up patterns giving
           the same final state with two identical particles interchanged..
          }
  \label{fig:Lundstring}
\end{figure}

In the Lund string model, the longitudinal break-up of the color string is governed by the area law.
The matrix element to get a final state depends on the area $A$:
   ${\cal M}=\exp\left[(\imath\kappa-b/2)A\right]$,
where $\kappa$ is the string tension and $b$ is a decay constant with values
$\kappa\approx1$\,\GeV/fm and $b\approx0.3$\,\GeV/fm.
Consider the string break-up illustrated by the solid line in Fig.~\ref{fig:Lundstring},
which spans the area $A$.
Suppose that the particles 1 and 2 are identical. The final state produced by interchanging them
would be produced by the string break-up with area $A+\Delta A$.
Transverse momentum arises via a tunneling mechanism, which is also related to $b$.
To incorporate BEC in the string model, the probability of a final state should be taken as
the square of the sum of the matrix elements corresponding to the areas of all the permutations of
identical bosons \cite{AH86,AR98,AR98a}.
This model results in BEC, including genuine 3-particle BEC.  It predicts that the longitudinal radius
is greater than the transverse radius and that the radius for neutral pions is smaller than that for charged
pions.
In such a model there is no mechanism for creating correlations between particles from different strings.
 
The model has been incorporated in Monte Carlo for a \Pq\Paq\ string.
While the formalism to do so for the more realistic case of a string with multiple gluons has been worked
out \cite{AMS01}, a successful Monte Carlo implementation has, unfortunately, so far proved elusive.

\section{Conclusions}
We have seen that the study of BEC in \Pee\ presents a number of problems, both experimental and
theoretical.  Consequently, values obtained for parameters vary considerably among experiments, even when
the same parametrization is used.  Nevertheless, certain features are clear:  BEC, both 2-particle and
genuine 3-particle,  exist; they seem independent of center-of-mass energy;  the source shape is somewhat
elongated in the jet direction; and the (Fermi-Dirac) radius for baryons is smaller than the radius for
mesons.
Experimentally, it is not clear whether the radius for neutral pions is smaller than that for charged pions.
BEC in W decay is the same as in light-quark Z decay. The data are compatible with no inter-W BEC.
 
The implementation of BEC in the Lund string model appears consistent with these experimental findings.
However, the experimental evidence that pion production is completely incoherent seems at odds with the
coherent addition of amplitudes in the model.

\bibliographystyle{aipproc}   
 
\bibliography{bec}
 
\IfFileExists{\jobname.bbl}{}
 {\typeout{}
  \typeout{******************************************} :74
  \typeout{** Please run "bibtex \jobname" to optain}
  \typeout{** the bibliography and then re-run LaTeX}
  \typeout{** twice to fix the references!}
  \typeout{******************************************}
  \typeout{}
 }
 
\end{document}